# A Central Disulfide Junction Drives Transient Network Formation in Elastin-Like Polypeptides, Enabling Low-Concentration Hydrogels

*Tingting Zhang,[1] Jean-François Le Meins,[1] Jean-Paul Chapel,[2] Saron Catak,[3] Guillaume Goudounet,[1] Olivier Sandre,[1] Nadia Mahmoudi,[1] Christophe Schatz[1,*] and Bertrand Garbay[1,*]*

[1] Laboratoire de Chimie des Polymères Organiques (LCPO), UMR 5629, Bordeaux University, CNRS, Bordeaux INP, F-33600, Pessac, France

[2] Centre de Recherche Paul Pascal (CRPP), UMR 5031, Bordeaux University, CNRS, F-33600, Pessac, France

[3] Department of chemistry, Bogazici University, Bebek 34342, Istanbul, Turkey

ABSTRACT

The self-assembly of associative triblock copolymers composed of a central hydrophilic elastin-like polypeptide (ELP) block and short fatty acid end groups (C16) was investigated in aqueous solution. In one system, the ELP contains 80 pentapeptide units (C16-80-C16), whereas in the other two C16-ELP40 chains were oxidatively coupled through their terminal cysteine residues to form a central disulfide bond, yielding C16-$(40)_2$-C16. Despite their nearly identical molecular weights and compositions, the two polymers exhibit markedly different self-assembly behaviors. C16-80-C16 forms large hydrophobic aggregates that remain kinetically trapped and do not develop a dynamically connected network. In contrast, C16-$(40)_2$-C16 forms very small associative nodes with an aggregation number of only ~3 chains. These nodes coexist with larger clusters and become dynamically interconnected at higher concentrations, leading to transparent hydrogels at concentrations as low as 2.5 wt.%. Oscillatory rheology reveals a transient Maxwell network governed by a single relaxation process associated with the reversible association of the C16 end groups. SAXS, light scattering, cryo-TEM, and molecular modeling consistently support a model in which the central disulfide junction promotes transient network formation.

Polymer gels from associative polymers are versatile materials with applications in drug delivery, self-healing, and rheological properties modification.[1–4] Among these, telechelic ABA polymers, with hydrophobic A blocks and hydrophilic B blocks, can form dynamic networks with tunable viscoelasticity in aqueous media. At low concentrations, they self-assemble into 'flower-like' micelles with hydrophobic cores (A blocks) and coronas of flexible loops, or 'petals' (B blocks). As concentration increases, loops begin to dissociate, forming bridges that link micelles into cluster-type associations.[5] Upon reaching a critical concentration, these bridges percolate into a space-spanning transient network of interconnected micelles. Gel-like behavior emerges when bridge lifetimes exceed the timescale of mechanical solicitation.[3] A well-known example of transient networks forming gels at low concentrations, typically from about 1 wt.% are hydrophobically modified ethoxylated urethanes (HEURs), consisting of poly(ethylene oxide) chains (5,000–40,000 g/mol) end-capped with short alkyl groups (C6–C20) through an urethane bond.[6–13] In these systems, stress relaxation follows a simple Maxwell model, governed by the escape of hydrophobic blocks from micelle cores.[8]

With their simple sequences, lack of stable tertiary structure, and precise control over length and composition, intrinsically disordered proteins (IDPs) can serve as well-defined model polymer systems for fundamental studies of chain conformations, phase behavior, and network formation.[14,15] Among these, elastin-like polypeptides (ELPs) are inspired by the disordered regions of tropoelastin and consist of repeats of the pentapeptide VPGXG, where X is a guest residue. ELPs undergo a thermally triggered phase transition at a sequence-dependent cloud point temperature ($T_{cp}$), also referred as the lower critical solution temperature (LCST).[16–18] This property has been exploited in applications such as protein purification, affinity capture, immunoassays, and drug delivery.[19,20] ELPs can also form hydrogels, which serve as scaffolds for tissue engineering, drug release platforms, and other biomedical applications. Most ELP hydrogels are formed *via* chemical, enzymatic, or γ-irradiation cross-linking, yielding stable covalent networks.[21–26] Physical gels from telechelic ELPs with a high-$T_{cp}$ central block and low-$T_{cp}$ end blocks, have also been reported,[27–31] but none form HEUR-type transient networks with reversibly interconnected micelles. Instead, their rheology reflects a jammed micelle state, consistent with the high polymer concentrations (> 20 wt.%) needed for gel formation.[27–29]

Given this, we hypothesized that the ELP end-blocks used in previous studies were likely too long to enable dynamic bridging between micelles. Inspired by HEUR polymers, we therefore designed ABA triblock copolymers comprising a central ELP block flanked by C16 palmitic acids at each end. We further reasoned that the conformation of the central hydrophilic block is equally critical for network dynamics, a parameter largely overlooked in HEUR systems. Inspired by nature, we considered molecular hinges in proteins: in immunoglobulin G (IgG), disulfide bonds in the hinge region link Fab and Fc domains, stabilizing the overall structure while permitting rotation and bending of the Fab arms to accommodate diverse epitopes.[32] Similarly, in certain hydrolases, transport proteins, or thioredoxin-like folds, an S–S linkage connects mobile helices or loops, acting as a controlled axis of motion.[33] By analogy, we propose that introducing a disulfide bond into the central ELP block could function as a covalent hinge that guides self-assembly and enhances dynamics, thereby promoting the formation of a transient network that enables gelation at low concentrations.

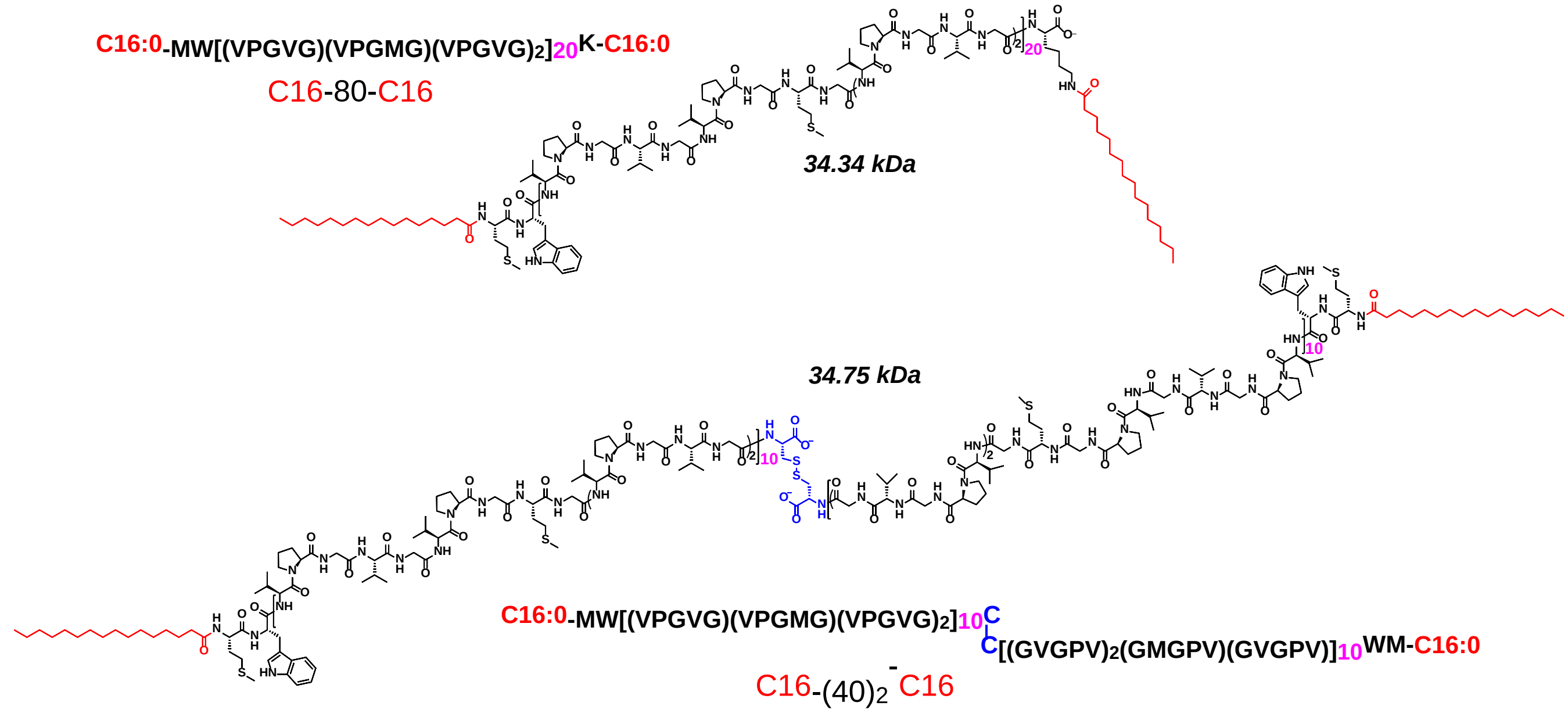

**Figure 1.** Chemical structures of palmitoylated ELP (80) (top) and ELP $(40)_2$ (bottom). Polypeptide sequences are shown in the one-letter amino acid code. The ELP backbone is depicted in black, with cysteine residues highlighted in blue and palmitic acid groups in red.

Two ELP constructs with similar compositions of approximately 400 amino acids were considered; their only difference is the presence of a central disulfide bond within the ELP block of the second construct. The first construct ($ELP_{80}$) consists of a single block of 80 pentapeptide repeats terminated by a lysine residue, $MW[(VPGVG)(VPGMG)(VPGVG)_2]_{20}K$. The second ELP $(40)_2$ comprises two identical 40-repeat sequences terminated by cysteine residues, $MW[(VPGVG)(VPGMG)(VPGVG)_2]_{10}C$, enabling disulfide linkage. Both constructs were produced recombinantly in *E. coli* and purified by inverse transition cycling (Supporting Information, Section 1.4). Lysine and cysteine codons were introduced by site-directed mutagenesis (Supporting Information, Section 1.3).[34,35] Oxidation of ELP40 with $H_2O_2$ under mild conditions yielded the disulfide-bridged $ELP(40)_2$ in approximately 95% purity, together with ~5% residual monomeric ELP40, as determined by SDS-PAGE densitometry (Supporting

Information, Section 1.5; Figures S1–S2). The purified ELP80 and ELP$(40)_2$ were subsequently acylated on primary amines with palmitic acid (C16:0) as described previously, and analyzed by RP-HPLC (Supporting Information, Section 1.6; Figure S3).[36] In the absence of reducing agents, the disulfide bond is expected to remain stable throughout sample storage and characterization.[37]

First, the solution behavior of the ELPs was examined under dilute conditions (2.5 mg.mL$^{-1}$). The non-acylated ELP80 and ELP$(40)_2$ showed sharp increases in turbidity upon heating, characteristic of the LCST transition (Figure 2, top). Their cloud points ($T_{cp}$) were 28°C and 30°C, respectively, indicating slightly greater hydration for ELP$(40)_2$. The transition corresponded to a reversible liquid–liquid phase separation (coacervation), confirmed by optical microscopy (Figure S4 and Videos S1-S4), with minor hysteresis likely due to the low cooling rate applied. Importantly, the cloud point temperature is highly tunable and can be shifted upward or downward by altering the ELP sequence, allowing precise control over the onset of coacervation.[17,18] For the palmitoylated ELPs, three notable effects were observed (Figure 2, middle). First, measurable optical density (OD) values appeared at 10 °C, indicating turbidity from small scattering structures forming well below the $T_{cp}$. Second, the $T_{cp}$ was much lower, reflecting reduced solubility due to alkyl interactions, as previously reported for hydrophobically modified telechelic poly(N-isopropylacrylamide) (PNIPAM)[38,39] and modified ELPs.[36,40,41]. Third, the increase in OD above $T_{cp}$ was much smaller than for unmodified ELPs, suggesting that the self-assembled structures resist aggregation at elevated temperatures.

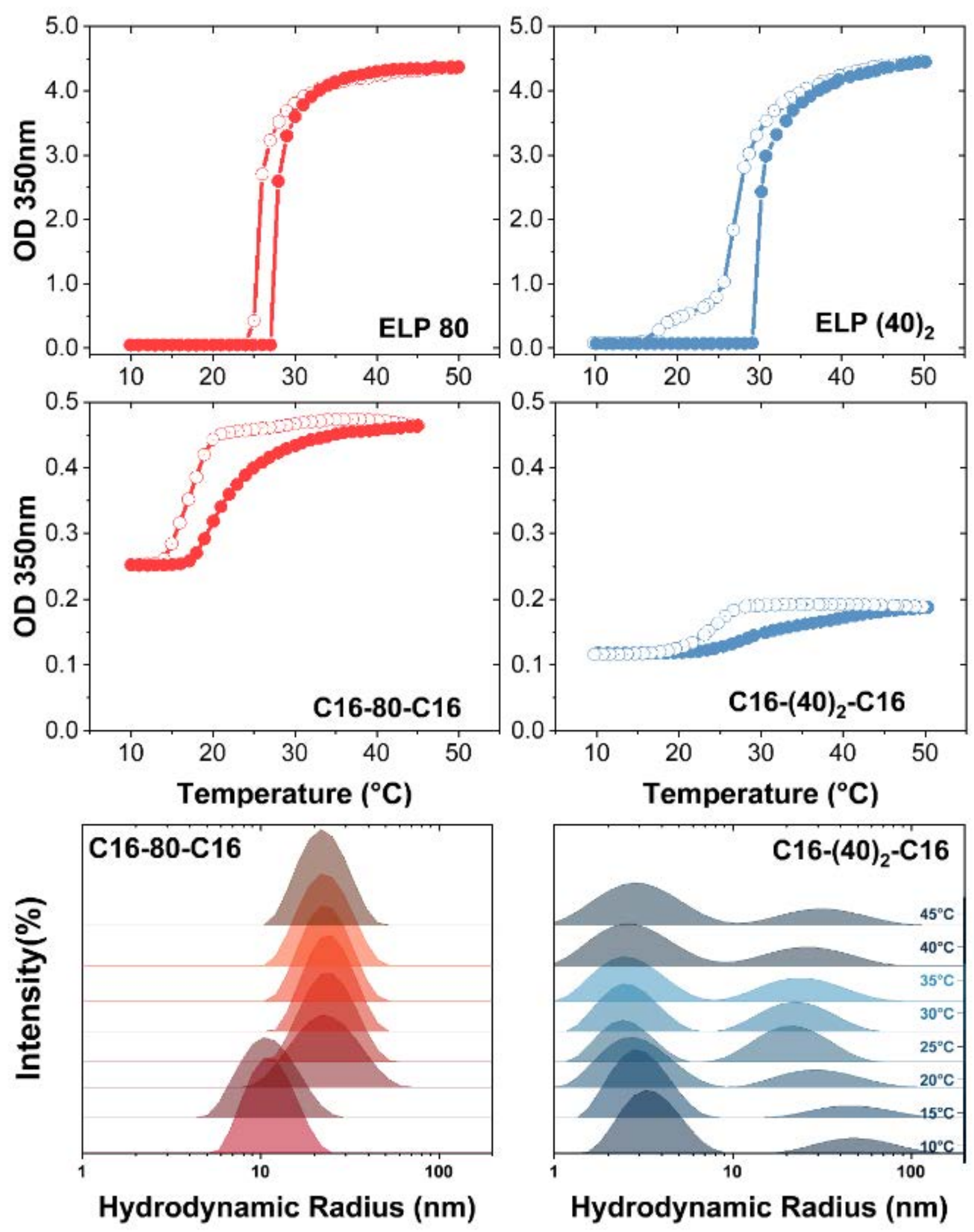

**Figure 2.** Solution behavior of ELPs and palmitoylated ELPs at 25 mg mL$^{-1}$ in water. (*top*): Optical density (OD) at $\lambda$ = 350 nm during heating (filled symbols) and cooling (empty symbols) ramps (1°C.min$^{-1}$) for unmodified ELP80 and ELP$(40)_2$). (*middle*): Corresponding OD profiles for palmitoylated C16-80-C16 and C16-$(40)_2$-C16. (*bottom*): DLS of palmitoylated ELPs, showing intensity-weighted size distributions for C16-80-C16 and C16-$(40)_2$-C16 between 10°C and 45°C.

Dynamic light scattering (DLS) at the same concentration revealed that C16-80-C16 formed a single population with hydrodynamic radius ($R_H$) ≈ 12 nm at 10°C, consistent with a micellar morphology (Figure 2, bottom). Upon heating to $T_{cp}$ ≈ 20°C, $R_H$ increased to 25 nm and remained stable, suggesting formation of mesoglobules, which are compact aggregates formed by collapsed chains above the demixing point, as reported for PNIPAM[42,43] and PNIPAM copolymers.[38,44] In contrast, C16-$(40)_2$-C16, with a central disulfide bond, displayed two populations: small particles ($R_H \approx 3.7$ nm at 10°C, decreasing slightly to 3.1 nm at 40°C) and larger particles ($R_H > 25$ nm) persisting across 10–40 °C (Figure 2, bottom). Such a coexistence of two populations suggests the presence of small dynamic associative aggregates in equilibrium with larger clusters. SAXS analysis of C16–80–C16 and C16–$(40)_2$–C16 supports the DLS results, indicating that C16–$(40)_2$–C16 forms markedly smaller associative aggregates whose characteristic dimensions remain largely unchanged over the investigated temperature range (Figure S9).

Static light scattering (SLS) measurements were performed over a polymer concentration range extending from 1 to 15 mg $mL^{-1}$ to further characterize the small associative aggregates identified by DLS. Throughout this concentration range, the hydrodynamic radius of the primary aggregates remained essentially unchanged, whereas the relative contribution of the larger clusters progressively increased (Figure S6). Such behavior is characteristic of closed associative aggregates in equilibrium with larger and more open cluster-like structure. Taking this concentration-dependent clustering into account, the scattering contribution of the primary aggregates could be estimated and analyzed separately. This yielded a weight-average molar mass (Mw) of $(1.12 \pm 0.20) \times 10^5$ g $mol^{-1}$, corresponding to an aggregation number (Nagg) of 3.1, together with a small positive second virial coefficient ($A_{2,z} = 5.5 \times 10^{-5}$ mol mL $g^{-2}$), indicating good solvent conditions (Figure S7). This aggregation number is substantially lower than that reported for the closely related telechelic polymer C16–PEO (30 kDa)–C16, which possesses the same C16 end groups and a hydrophilic block of comparable molar mass and typically forms aggregates containing 10–11 chains.[10,45] Despite this lower aggregation number, the concentration-dependent behavior remains similar to that reported for a wide range of associative polymers, with small associative aggregates coexisting with larger clusters whose relative contribution increases with concentration.[5,9,10,45–47]

Taken together, these results indicate that the primary associating species are very small hydrophobic aggregates rather than conventional polymer micelles. In telechelic associative polymers, flower-like micelles are generally regarded as the elementary self-assembled structures. However, their unambiguous identification remains difficult, as scattering experiments alone often cannot discriminate between flower-like and other associative morphologies.[48] The very low aggregation number observed here (Nagg ≈ 3) suggests that the primary species are more appropriately described as small associative or hydrophobic nodes formed by only a few chains. This low aggregation number is likely related to the specific conformational features introduced by the central disulfide junction. Another contributing factor may arise from the two carboxylate groups flanking the central disulfide bond (Figure 1). As commonly observed in charged micellar systems, electrostatic repulsion tends to reduce the aggregation number, as described by classical surfactant packing theory, while also favoring more dynamic association–dissociation processes within the associative nodes. Regarding their morphology, these nodes may be related to the branched or comb-like associative structures previously proposed for telechelic polymers.[49] At this stage, the precise morphology of the elementary associative species is probably of secondary importance. The key question is whether these nodes remain isolated or become dynamically interconnected at higher concentrations.

Rheological measurements were therefore performed under more concentrated conditions to determine whether the associative nodes identified by scattering become interconnected into a transient network at higher concentrations. Such dynamic connectivity can be assessed directly from the viscoelastic response. ELPs were dissolved in water at [illegible] °C for [illegible] h at 5 wt%, yielding hydrogels for both C16-80-C16 and C16-(40)$_2$-C16 (Figure 3). The C16-80-C16 hydrogel appeared whitish and flowed readily upon vial inversion, reflecting fast internal dynamics. In contrast, the C16-(40)$_2$-C16 hydrogel was transparent and remained immobile even after prolonged storage.

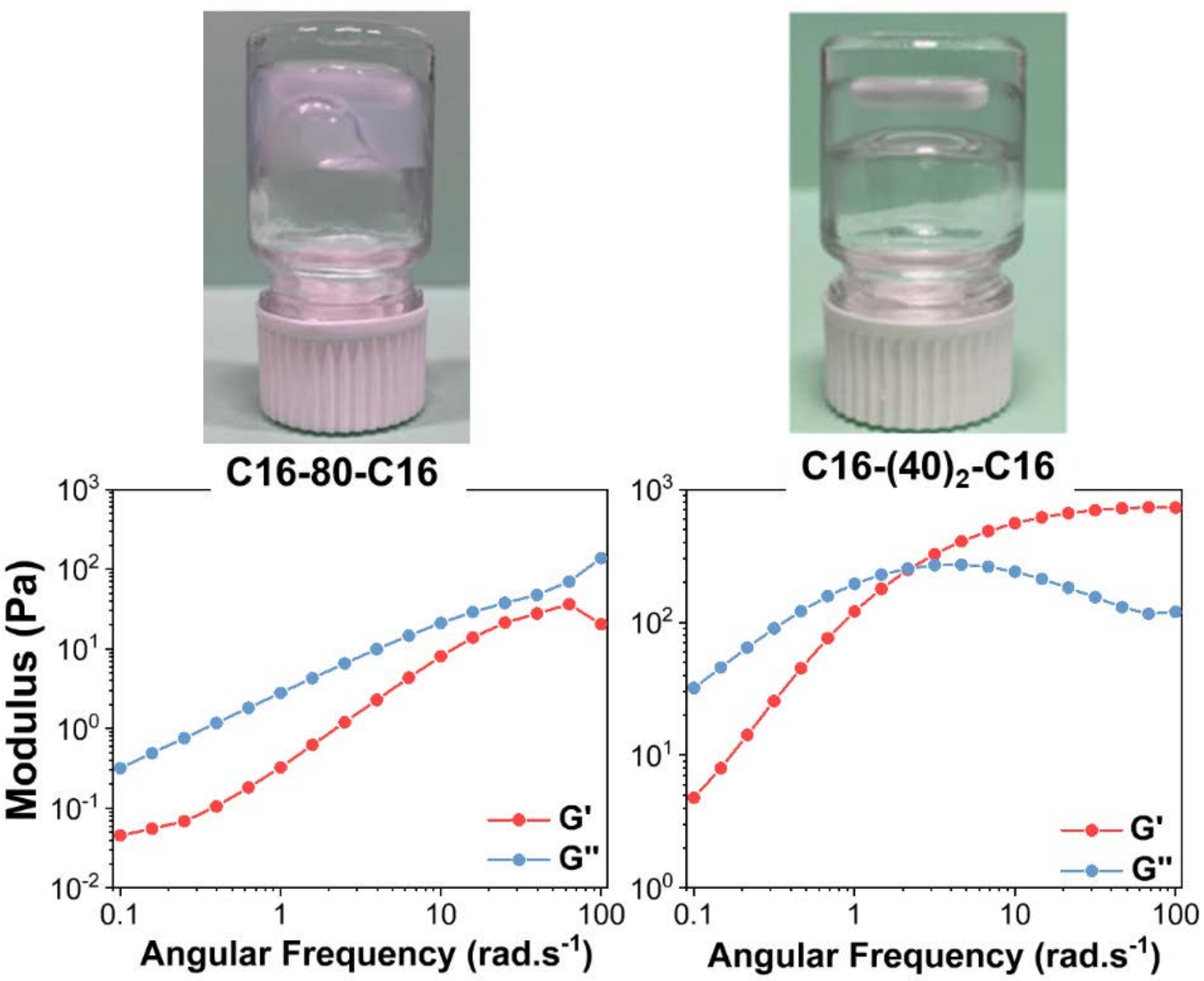

**Figure 3.** Visual appearance and oscillatory rheology of C16-80-C16 (left) and C16-(40)$_2$-C16 (right) gels at 5 wt.% and 10 °C. Storage modulus (G') and loss modulus (G'') were obtained during a frequency sweep at a fixed strain amplitude of [illegible]%.

Rheological characterization confirmed these observations. The C16-80-C16 hydrogel exhibited viscoelastic fluid characteristics, with viscous behavior dominating across the entire frequency range studied. No crossover between the storage modulus (*G*') and the loss modulus (*G*'') was observed up to 100 rad s$^{-1}$. At low frequencies, *G*'' scaled approximately with $\omega$, while *G*' followed an $\omega^2$ dependence, consistent with the terminal flow regime where all structural relaxation processes have been completed. The minor plateau for *G*' at very low frequencies was attributed to measurement artifacts. In contrast, the C16-$(40)_2$-C16 hydrogel displayed the characteristic rheological response of a Maxwell fluid. The system exhibited predominantly elastic behavior, with a storage plateau modulus ($G_0 \approx 800$ Pa) above 10 rad.s$^{-1}$, and transitioned to a terminal flow regime at lower frequencies. This transition was marked by a *G*'-*G*'' crossover near $G_0/2$, indicating a relatively narrow relaxation time distribution ($\tau \approx 0.5$ s at 10°C). Such behavior is characteristic of a transient network formed through reversible connections between associative nodes.[8,10,11] The relaxation time, governed by the dynamics of the C16 end groups within the nodes, agrees with values reported at 20°C for associative polymers bearing the same hydrophobic end groups,[7,10] after temperature correction using the Arrhenius relation which is obeyed for such systems ($E_a \approx 60$ kJ.mol$^{-1}$).[8,50] The reversible association of C16 end groups imparts self-healing properties, allowing the network to reorganize after mechanical disruption (Video S5). Stable hydrogels could be formed at concentrations as low as 2.5 wt.% (Figure S10), with moduli comparable to chemically crosslinked ELP hydrogels (100–1000 Pa).[51–55] Finally, it should be noted that the C16-$(40)_2$-C16 sample contains approximately 5% residual monochelic ELP40 (Figure S1). As reported for associative telechelic polymers, monochelic impurities reduce the density of elastically active bridges, leading to lower values of the plateau storage modulus ($G_0$); this effect increases with their concentration.[56] Under the present conditions, however, only a limited quantitative impact on the rheological properties is expected.

Upon heating, the C16-$(40)_2$-C16 gel exhibited a gradual decrease in *G*' and a more moderate decrease in *G*'', with a cross-over near 23°C indicating the loss of gel properties (Figure S11). As previously observed in dilute solution, heating caused a small but significant increase in OD and light scattering intensity, consistent with partial dehydration of the central ELP block (Figure 2 and Figure S5). This destabilization weakened the dynamic connectivity between associative nodes, leading to a sharp drop in viscoelasticity and network collapse. Upon cooling, $G'$ recovered and exceeded $G''$ below 15 °C, showing complete gel reformation. The reversible response demonstrates that the system remains in thermodynamic equilibrium throughout the cycle.

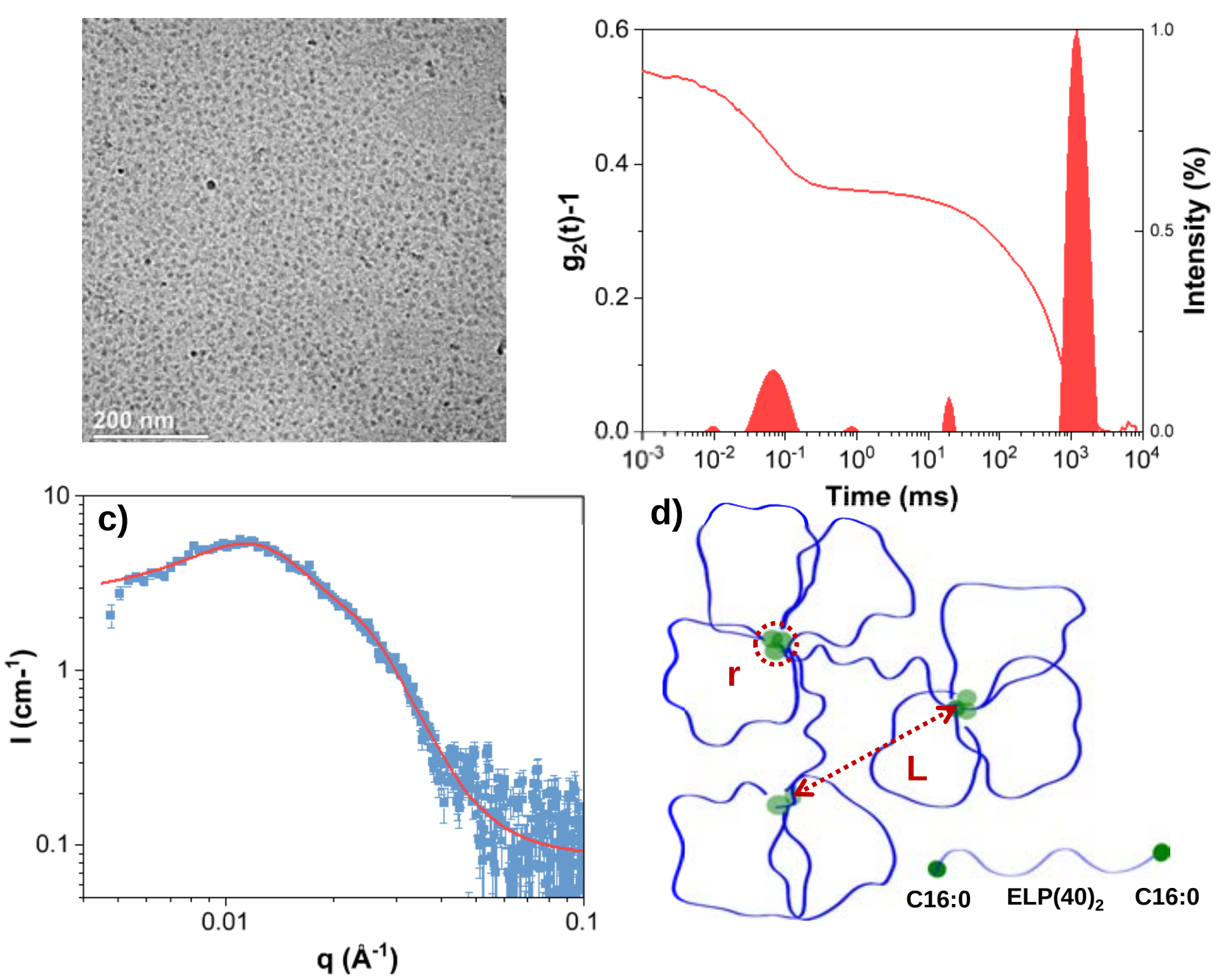

**Figure 4**. Characterization of the C16-(40)$_2$-C16 gel at 10 °C. a) cryo-TEM (5 wt%). b) DLS at 10 (2.5 wt.%). c) SAXS (5 wt.%). d) Characteristics length scales in the gel (r, L) determined from SAXS data.

Additional characterization of the C16-(40)$_2$-C16 hydrogel at 10 °C was performed using cryo-TEM, DLS, and SAXS. Cryo-TEM revealed well-defined associative structures with sizes close to those observed under dilute conditions ($R$ = 4.2 ± 1.7 nm; Figure 4a and Figure S12), supporting a closed-association model that remains unchanged with concentration. DLS analysis conducted just above the sol–gel transition (2.5 wt%), where the system remains ergodic, showed two distinct relaxation modes (Figure 4b). The fast mode ($\tau$ = 0.067 ms) which corresponds to a hydrodynamic radius of 3.7 ± 0.4 nm is still consistent at this concentration with the cooperative diffusion of the primary associative aggregates. The slower mode ($\tau$ = 1.25 s) reflects the dynamics of large-scale structures ($r \approx 70$ nm) commonly observed in associative telechelic polymer systems, but whose origin is not obvious (e.g., spatial heterogeneities or network rearrangements).[5,13,57–59] Small-angle X-ray scattering (SAXS) of the gel prepared under similar conditions was analyzed using a polydisperse sphere form factor combined with a hard-sphere structure factor (Figure 4c and Figure S13). The fitted particle radius ($r$ = 4.9 ± 1.3 nm) was slightly larger than that from light scattering, likely due to contrast differences between X-ray and optical techniques. The associative nodes interact primarily through transient bridging of the central hydrophilic ELP blocks rather than simple excluded-volume effects, yielding an apparent hard-sphere radius of $R_{HS}$ = 24 nm, corresponding to an average center-to-center distance of $L$ = 48 nm (Figure 4d). This distance corresponds to approximately one-third of the contour length of the ELP chain ($L_c \approx 146$ nm, assuming 0.365 nm per amino-acid residue), indicating that such bridging is geometrically accessible. This spacing is also in good agreement with the value inferred from the correlation peak ($q^*$), which gives $L = 2\pi/q^* \approx 52$ nm (Figure 4c) after excluding the form factor contribution (Figure S13). The volume fraction of the nodes obtained from the structure factor, $\phi$ = 0.13, is consistent with values reported for systems exhibiting comparable intermicellar spacing,[60] confirming that gelation is due to bridging and not to jamming ($\phi$ = 0.64).[61]

The self-assembly behaviors of C16-$(40)_2$-C16 differs markedly from that of C16-80-C16. While C16-80-C16 forms relatively stable associative aggregates that evolve into mesoglobules upon heating, C16-$(40)_2$-C16 forms small associative nodes (Nagg ≈ 3) that become dynamically interconnected into a transient network. Since both copolymers bear identical C16 end groups, this difference is therefore likely to originate from the architecture of the central ELP block. Consistent with this interpretation, SAXS measurements performed on the non-acylated ELP80 and $ELP(40)_2$ precursors revealed a smaller radius of gyration for $ELP(40)_2$ ($R_G$ = 6.1 nm versus 6.9 nm), suggesting a more flexible chain conformation (Figure S8). Quantum-mechanical simulations were therefore performed to investigate the molecular basis of this difference.[62] A methyl-truncated polymer chain (PGVGCCGVGP) was used to mimic the central part of the copolymer containing the disulfide bridge between the two cysteine (C) residues. Implicit solvent effects were included in the calculations. The optimized geometries revealed characteristic bond (C–S–S ≈ 105°) and dihedral (C–S–S–C ≈ 90°) angles, which are a direct consequence of the participation of sulfur d-orbitals (notably $d_{x^2-y^2}$ and $d_{z^2}$) in bonding.[71] As shown in Figure 5, the dihedral angle gives rise to an almost orthogonal arrangement that acts as a molecular hinge, restricting the conformational freedom of the central block. Consequently, the chain can adopt two dominant geometries—one almost linear and the other bent. This conformational bistability provides a plausible molecular basis for a dynamic equilibrium between discrete molecular states: in the linear configuration, the hydrophobic termini are exposed, favoring intermolecular bridging and node formation, while in the bent state, these termini approach each other, promoting intramolecular association. This interpretation should, however, be considered in light of the fact that the molecular modeling was deliberately restricted to a truncated model of the central junction in order to isolate its intrinsic conformational properties. Although the model includes the terminal carboxylate groups, it describes only an isolated molecular fragment. More generally, additional physicochemical factors, including hydration of the ELP block, micellar packing, hydrophobic collapse, counterion screening, and collective thermal fluctuations, are also expected to contribute to the observed behavior.

a)

b)

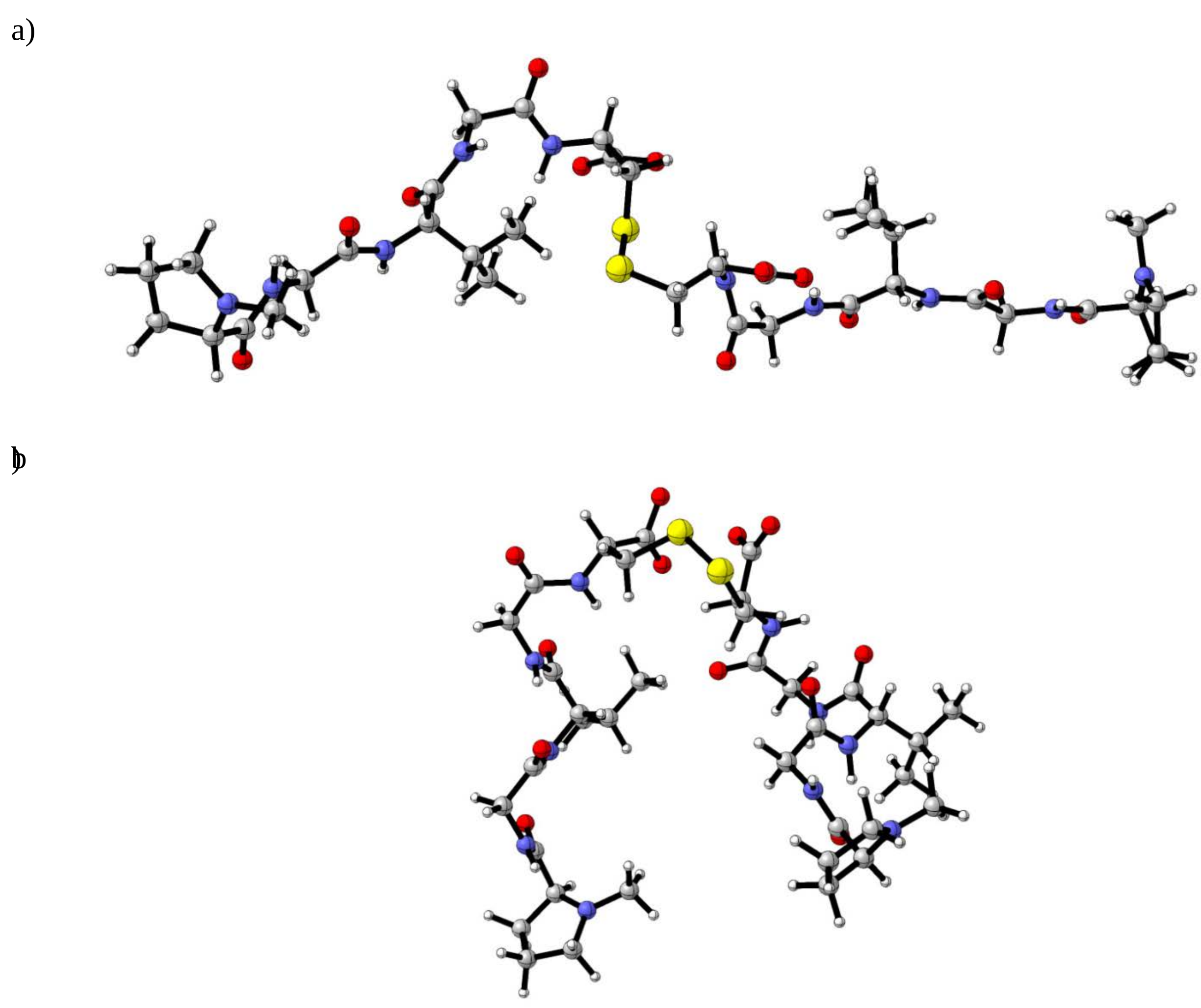

**Figure 5.** DFT optimized conformations of methyl-truncated PGVG**CC**GVGP central peptide (C: cysteine). a) linear conformation, b) convected conformation

Overall, these results demonstrate that oxidative coupling of two monothiolic ELP precursors through their terminal cysteine residues generates a telechelic ELP with a central disulfide junction that promotes transient network formation, enabling equilibrium hydrogels to form at remarkably low polymer concentrations. Similar strategies are found in biological systems, where proteins exploit bistable conformations to regulate function, from enzymes that alternate between open and closed states to signaling proteins that toggle between active and inactive forms.[64,65] An important next step will now be to investigate the broader role of the central junction itself. Inspired by the spacer chemistry widely used in bioconjugation, alternative junctions based on amide, triazole, thioether, PEG or neutral linkers, as well as charged spacers, could be explored to determine how the chemical nature and conformational properties of the central junction govern self-assembly dynamics and whether kinetically trapped associative systems can be converted into dynamic transient networks.

**Supporting Information**

Materials and methods, synthesis and characterization of ELPs, supplementary characterization data of hydrogels including light scattering, rheology and microscopy (Figures S1-S10, Videos S1-S5) and additional references.

**Corresponding Author**

christophe.schatz@bordeaux-inp.fr, bertrand.garbay@bordeaux-inp.fr

**Author Contributions**

The manuscript was written through contributions of all authors. All authors have given approval to the final version of the manuscript.

**Funding source**

This work was funded China Scholarship Council and a Université de Bordeaux grant (UB-CSC 2018).

**Acknowledgements**

E. Garanger and S. Lecommandoux are gratefully acknowledged for their constant support throughout this study. Jean-Michel Guinier from the Institut de Minéralogie, de Physique des Matériaux et de Cosmochimie (Paris, France) is thanked for Cryo-TEM imaging.

REFERENCES

(1) Winnik, M. A.; Yekta, A. Associative Polymers in Aqueous Solution. *Current Opinion in Colloid & Interface Science* **1997**, *2* (4), 424–436. https://doi.org/10.1016/S1359-0294(97)80088-X.
(2) Rubinstein, M.; Dobrynin, A. V. Associations Leading to Formation of Reversible Networks and Gels. *Current Opinion in Colloid & Interface Science* **1999**, *4* (1), 83–87. https://doi.org/10.1016/S1359-0294(99)00013-8.
(3) Chassenieux, C.; Nicolai, T.; Benyahia, L. Rheology of Associative Polymer Solutions. *Current Opinion in Colloid & Interface Science* **2011**, *16* (1), 18–26. https://doi.org/10.1016/j.cocis.2010.07.007.
(4) Lund, R.; Willner, L.; Holderer, O. Structure and Chain Dynamics of Self-Healing Telechelic Polymer Networks. *Macromolecules* **2025**, *58* (18), 9754–9762. https://doi.org/10.1021/acs.macromol.5c01216.
(5) Alami, E.; Almgren, M.; Brown, W.; François, J. Aggregation of Hydrophobically End-Capped Poly(Ethylene Oxide) in Aqueous Solutions. Fluorescence and Light-Scattering Studies. *Macromolecules* **1996**, *29* (6), 2229–2243. https://doi.org/10.1021/ma951174h.
(6) Tanaka, F.; Edwards, S. F. Viscoelastic Properties of Physically Crosslinked Networks. *Journal of Non-Newtonian Fluid Mechanics* **1992**, *43* (2–3), 247–271. https://doi.org/10.1016/0377-0257(92)80027-U.
(7) Fonnum, G.; Bakke, J.; Hansen, F. K. Associative Thickeners. Part I: Synthesis, Rheology and Aggregation Behavior. *Colloid Polym Sci* **1993**, *271* (4), 380–389. https://doi.org/10.1007/BF00657419.
(8) Annable, T.; Buscall, R.; Ettelaie, R.; Whittlestone, D. The Rheology of Solutions of Associating Polymers: Comparison of Experimental Behavior with Transient Network Theory. *Journal of Rheology* **1993**, *37* (4), 695–726. https://doi.org/10.1122/1.550391.
(9) Semenov, A. N.; Joanny, J.-F.; Khokhlov, A. R. Associating Polymers: Equilibrium and Linear Viscoelasticity. *Macromolecules* **1995**, *28* (4), 1066–1075. https://doi.org/10.1021/ma00108a038.
(10) Xu, B.; Yekta, A.; Li, L.; Masoumi, Z.; Winnik, M. A. The Functionality of Associative Polymer Networks: The Association Behavior of Hydrophobically Modified Urethane-Ethoxylate (HEUR) Associative Polymers in Aqueous Solution. *Colloids and Surfaces A: Physicochemical and Engineering Aspects* **1996**, *112* (2–3), 239–250. https://doi.org/10.1016/0927-7757(96)03558-3.
(11) Tam, K. C.; Jenkins, R. D.; Winnik, M. A.; Bassett, D. R. A Structural Model of Hydrophobically Modified Urethane−Ethoxylate (HEUR) Associative Polymers in Shear Flows. *Macromolecules* **1998**, *31* (13), 4149–4159. https://doi.org/10.1021/ma980148r.
(12) Quienne, B.; Pinaud, J.; Robin, J.-J.; Caillol, S. From Architectures to Cutting-Edge Properties, the Blooming World of Hydrophobically Modified Ethoxylated Urethanes (HEURs). *Macromolecules* **2020**, *53* (16), 6754–6766. https://doi.org/10.1021/acs.macromol.0c01353.
(13) Katashima, T. Precise Analysis of Rheological Properties of Transient Network Using Model Materials. *Progress in Polymer Science* **2025**, *171*, 102042. https://doi.org/10.1016/j.progpolymsci.2025.102042.

(14) Garcia Quiroz, F.; Li, N. K.; Roberts, S.; Weber, P.; Dzuricky, M.; Weitzhandler, I.; Yingling, Y. G.; Chilkoti, A. Intrinsically Disordered Proteins Access a Range of Hysteretic Phase Separation Behaviors. *Sci. Adv.* **2019**, *5* (10), eaax5177. https://doi.org/10.1126/sciadv.aax5177.
(15) Chakraborty, S.; Morozova, T. I.; Barrat, J.-L. Intrinsically Disordered Proteins Can Behave as Different Polymers across Their Conformational Ensemble. *J. Phys. Chem. B* **2025**, *129* (9), 2359–2369. https://doi.org/10.1021/acs.jpcb.4c07020.
(16) Urry, D. W.; Haynes, B.; Harris, R. D. Temperature Dependence of Length of Elastin and Its Polypentapeptide. *Biochemical and Biophysical Research Communications* **1986**, *141* (2), 749–755. https://doi.org/10.1016/S0006-291X(86)80236-4.
(17) Meyer, D. E.; Chilkoti, A. Quantification of the Effects of Chain Length and Concentration on the Thermal Behavior of Elastin-like Polypeptides. *Biomacromolecules* **2004**, *5* (3), 846–851. https://doi.org/10.1021/bm034215n.
(18) Aladini, F.; Araman, C.; Becker, C. F. W. Chemical Synthesis and Characterization of Elastin-like Polypeptides (ELPs) with Variable Guest Residues. *Journal of Peptide Science* **2016**, *22* (5), 334–342. https://doi.org/10.1002/psc.2871.
(19) Varanko, A. K.; Su, J. C.; Chilkoti, A. Elastin-Like Polypeptides for Biomedical Applications. *Annu. Rev. Biomed. Eng.* **2020**, *22* (1), 343–369. https://doi.org/10.1146/annurev-bioeng-092419-061127.
(20) Guo, Y.; Liu, S.; Jing, D.; Liu, N.; Luo, X. The Construction of Elastin-like Polypeptides and Their Applications in Drug Delivery System and Tissue Repair. *J Nanobiotechnol* **2023**, *21* (1), 418. https://doi.org/10.1186/s12951-023-02184-8.
(21) Lee, J.; Macosko, C. W.; Urry, D. W. Mechanical Properties of Cross-Linked Synthetic Elastomeric Polypentapeptides. *Macromolecules* **2001**, *34* (17), 5968–5974. https://doi.org/10.1021/ma0017844.
(22) Nagapudi, K.; Brinkman, W. T.; Leisen, J. E.; Huang, L.; McMillan, R. A.; Apkarian, R. P.; Conticello, V. P.; Chaikof, E. L. Photomediated Solid-State Cross-Linking of an Elastin−Mimetic Recombinant Protein Polymer. *Macromolecules* **2002**, *35* (5), 1730–1737. https://doi.org/10.1021/ma011429t.
(23) Trabbic-Carlson, K.; Setton, L. A.; Chilkoti, A. Swelling and Mechanical Behaviors of Chemically Cross-Linked Hydrogels of Elastin-like Polypeptides. *Biomacromolecules* **2003**, *4* (3), 572–580. https://doi.org/10.1021/bm025671z.
(24) Lim, D. W.; Nettles, D. L.; Setton, L. A.; Chilkoti, A. Rapid Cross-Linking of Elastin-like Polypeptides with (Hydroxymethyl)Phosphines in Aqueous Solution. *Biomacromolecules* **2007**, *8* (5), 1463–1470. https://doi.org/10.1021/bm061059m.
(25) Zhang, Y.; Avery, R. K.; Vallmajo-Martin, Q.; Assmann, A.; Vegh, A.; Memic, A.; Olsen, B. D.; Annabi, N.; Khademhosseini, A. A Highly Elastic and Rapidly Crosslinkable Elastin-Like Polypeptide-Based Hydrogel for Biomedical Applications. *Adv Funct Materials* **2015**, *25* (30), 4814–4826. https://doi.org/10.1002/adfm.201501489.
(26) Dos Santos, B. P.; Garbay, B.; Fenelon, M.; Rosselin, M.; Garanger, E.; Lecommandoux, S.; Oliveira, H.; Amédée, J. Development of a Cell-Free and Growth Factor-Free Hydrogel Capable of Inducing Angiogenesis and Innervation after Subcutaneous Implantation. *Acta Biomaterialia* **2019**, *99*, 154–167. https://doi.org/10.1016/j.actbio.2019.08.028.
(27) Wright, E. R.; McMillan, R. A.; Cooper, A.; Apkarian, R. P.; Conticello, V. P. Thermoplastic Elastomer Hydrogels via Self-Assembly of an Elastin-Mimetic Triblock Polypeptide. *Adv.*

*Funct. Mater.* **2002**, *12* (2), 149–154. https://doi.org/10.1002/1616-3028(20020201)12:2%3C149::AID-ADFM149%3E3.0.CO;2-N.

(28) Wright, E. R.; Conticello, V. P. Self-Assembly of Block Copolymers Derived from Elastin-Mimetic Polypeptide Sequences. *Advanced Drug Delivery Reviews* **2002**, *54* (8), 1057–1073. https://doi.org/10.1016/S0169-409X(02)00059-5.

(29) Nagapudi, K.; Brinkman, W. T.; Thomas, B. S.; Park, J. O.; Srinivasarao, M.; Wright, E.; Conticello, V. P.; Chaikof, E. L. Viscoelastic and Mechanical Behavior of Recombinant Protein Elastomers. *Biomaterials* **2005**, *26* (23), 4695–4706. https://doi.org/10.1016/j.biomaterials.2004.11.027.

(30) Sing, M. K.; Burghardt, W. R.; Olsen, B. D. Influence of End-Block Dynamics on Deformation Behavior of Thermoresponsive Elastin-like Polypeptide Hydrogels. *Macromolecules* **2018**, *51* (8), 2951–2960. https://doi.org/10.1021/acs.macromol.8b00002.

(31) Dai, M.; Goudounet, G.; Zhao, H.; Garbay, B.; Garanger, E.; Pecastaings, G.; Schultze, X.; Lecommandoux, S. Thermosensitive Hybrid Elastin-like Polypeptide-Based ABC Triblock Hydrogel. *Macromolecules* **2021**, *54* (1), 327–340. https://doi.org/10.1021/acs.macromol.0c01744.

(32) Horx, P.; Geyer, A. Comparing the Hinge-Type Mobility of Natural and Designed Intermolecular Bi-Disulfide Domains. *Frontiers in Chemistry* **2020**, *Volume 8-2020*.

(33) Eble, J. A. Allosteric Disulfide Bridges in Integrins: The Molecular Switches of Redox Regulation of Integrin-Mediated Cell Functions. *Antioxidants* **2025**, *14* (8). https://doi.org/10.3390/antiox14081005.

(34) Petitdemange, R.; Garanger, E.; Bataille, L.; Dieryck, W.; Bathany, K.; Garbay, B.; Deming, T. J.; Lecommandoux, S. Selective Tuning of Elastin-like Polypeptide Properties via Methionine Oxidation. *Biomacromolecules* **2017**, *18* (2), 544–550. https://doi.org/10.1021/acs.biomac.6b01696.

(35) Dai, M.; Georgilis, E.; Goudounet, G.; Garbay, B.; Pille, J.; Van Hest, J. C. M.; Schultze, X.; Garanger, E.; Lecommandoux, S. Refining the Design of Diblock Elastin-Like Polypeptides for Self-Assembly into Nanoparticles. *Polymers* **2021**, *13* (9), 1470. https://doi.org/10.3390/polym13091470.

(36) Zhang, T.; Peruch, F.; Weber, A.; Bathany, K.; Fauquignon, M.; Mutschler, A.; Schatz, C.; Garbay, B. Solution Behavior and Encapsulation Properties of Fatty Acid–Elastin-like Polypeptide Conjugates. *RSC Adv.* **2023**, *13* (3), 2190–2201. https://doi.org/10.1039/D2RA06603C.

(37) Gilbert, H. F. [2] Thiol/Disulfide Exchange Equilibria and Disulfidebond Stability. In *Methods in Enzymology*; Academic Press, 1995; Vol. 251, pp 8–28. https://doi.org/10.1016/0076-6879(95)51107-5.

(38) Kujawa, P.; Tanaka, F.; Winnik, F. M. Temperature-Dependent Properties of Telechelic Hydrophobically Modified Poly(N-Isopropylacrylamides) in Water:  Evidence from Light Scattering and Fluorescence Spectroscopy for the Formation of Stable Mesoglobules at Elevated Temperatures. *Macromolecules* **2006**, *39* (8), 3048–3055. https://doi.org/10.1021/ma0600254.

(39) Kujawa, P.; Segui, F.; Shaban, S.; Diab, C.; Okada, Y.; Tanaka, F.; Winnik, F. M. Impact of End-Group Association and Main-Chain Hydration on the Thermosensitive Properties of Hydrophobically Modified Telechelic Poly(N-Isopropylacrylamides) in Water. *Macromolecules* **2006**, *39* (1), 341–348. https://doi.org/10.1021/ma051876z.

(40) Ji, J.; Hossain, M. S.; Krueger, E. N.; Zhang, Z.; Nangia, S.; Carpentier, B.; Martel, M.; Nangia, S.; Mozhdehi, D. Lipidation Alters the Structure and Hydration of Myristoylated Intrinsically Disordered Proteins. *Biomacromolecules* **2023**, *24* (3), 1244–1257. https://doi.org/10.1021/acs.biomac.2c01309.
(41) Zhang, Z.; Lynch, C. J.; Huo, Y.; Chakraborty, S.; Cremer, P. S.; Mozhdehi, D. Modulating Phase Behavior in Fatty Acid-Modified Elastin-like Polypeptides (FAMEs): Insights into the Impact of Lipid Length on Thermodynamics and Kinetics of Phase Separation. *J. Am. Chem. Soc.* **2024**, *146* (8), 5383–5392. https://doi.org/10.1021/jacs.3c12791.
(42) Halperin, A.; Kröger, M.; Winnik, F. M. Poly(N-Isopropylacrylamide) Phase Diagrams: Fifty Years of Research. *Angewandte Chemie International Edition* **2015**, *54* (51), 15342–15367. https://doi.org/10.1002/anie.201506663.
(43) Kujawa, P.; Aseyev, V.; Tenhu, H.; Winnik, F. M. Temperature-Sensitive Properties of Poly( *N* -Isopropylacrylamide) Mesoglobules Formed in Dilute Aqueous Solutions Heated above Their Demixing Point. *Macromolecules* **2006**, *39* (22), 7686–7693. https://doi.org/10.1021/ma061604b.
(44) Timoshenko, E. G.; Basovsky, R.; Kuznetsov, Y. A. Micellesation vs Aggregation in Dilute Solutions of Amphiphilic Heteropolymers. *Colloids and Surfaces A: Physicochemical and Engineering Aspects* **2001**, *190* (1), 129–134. https://doi.org/10.1016/S0927-7757(01)00672-0.
(45) Yekta, A.; Xu, B.; Duhamel, J.; Adiwidjaja, H.; Winnik, M. A. Fluorescence Studies of Associating Polymers in Water: Determination of the Chain End Aggregation Number and a Model for the Association Process. *Macromolecules* **1995**, *28* (4), 956–966. https://doi.org/10.1021/ma00108a025.
(46) Yekta, A.; Duhamel, J.; Adiwidjaja, H.; Brochard, P.; Winnik, M. A. Association Structure of Telechelic Associative Thickeners in Water. *Langmuir* **1993**, *9* (4), 881–883. https://doi.org/10.1021/la00028a001.
(47) Alami, E.; Abrahmsén-Alami, S.; Vasilescu, M.; Almgren, M. A Comparison between Hydrophobically End-Capped Poly(Ethylene Oxide) with Ether and Urethane Bonds. *Journal of Colloid and Interface Science* **1997**, *193* (2), 152–162. https://doi.org/10.1006/jcis.1997.5043.
(48) Séréro, Y.; Aznar, R.; Porte, G.; Berret, J.-F.; Calvet, D.; Collet, A.; Viguier, M. Associating Polymers: From ``Flowers'' to Transient Networks. *Phys. Rev. Lett.* **1998**, *81* (25), 5584–5587. https://doi.org/10.1103/PhysRevLett.81.5584.
(49) Maechling-Strasser, C.; Clouet, F.; Francois, J. Hydrophobically End-Capped Polyethylene-Oxide Urethanes: 2. Modelling Their Association in Water. *Polymer* **1992**, *33* (5), 1021–1025. https://doi.org/10.1016/0032-3861(92)90018-R.
(50) Le Meins, J.-F.; Tassin, J.-F. Elastic Modulus and Relaxation Times in Telechelic Associating Polymers. *Colloid Polym Sci* **2003**, *281* (3), 283–287. https://doi.org/10.1007/s00396-002-0708-x.
(51) Xu, D.; Asai, D.; Chilkoti, A.; Craig, S. L. Rheological Properties of Cysteine-Containing Elastin-Like Polypeptide Solutions and Hydrogels. *Biomacromolecules* **2012**, *13* (8), 2315–2321. https://doi.org/10.1021/bm300760s.
(52) Cai, L.; Dinh, C. B.; Heilshorn, S. C. One-Pot Synthesis of Elastin-like Polypeptide Hydrogels with Grafted VEGF-Mimetic Peptides. *Biomater. Sci.* **2014**, *2* (5), 757–765. https://doi.org/10.1039/C3BM60293A.

(53) Wang, H.; Cai, L.; Paul, A.; Enejder, A.; Heilshorn, S. C. Hybrid Elastin-like Polypeptide–Polyethylene Glycol (ELP-PEG) Hydrogels with Improved Transparency and Independent Control of Matrix Mechanics and Cell Ligand Density. *Biomacromolecules* **2014**, *15* (9), 3421–3428. https://doi.org/10.1021/bm500969d.
(54) Contessotto, P.; Orbanić, D.; Da Costa, M.; Jin, C.; Owens, P.; Chantepie, S.; Chinello, C.; Newell, J.; Magni, F.; Papy-Garcia, D.; Karlsson, N. G.; Kilcoyne, M.; Dockery, P.; Rodríguez-Cabello, J. C.; Pandit, A. Elastin-like Recombinamers-Based Hydrogel Modulates Post-Ischemic Remodeling in a Non-Transmural Myocardial Infarction in Sheep. *Sci. Transl. Med.* **2021**, *13* (581), eaaz5380. https://doi.org/10.1126/scitranslmed.aaz5380.
(55) Wang, H.; Zhu, D.; Paul, A.; Cai, L.; Enejder, A.; Yang, F.; Heilshorn, S. C. Covalently Adaptable Elastin-Like Protein–Hyaluronic Acid (ELP–HA) Hybrid Hydrogels with Secondary Thermoresponsive Crosslinking for Injectable Stem Cell Delivery. *Advanced Functional Materials* **2017**, *27* (28), 1605609. https://doi.org/10.1002/adfm.201605609.
(56) Calvet, D.; Collet, A.; Viguier, M.; Berret, J.-F.; Séréro, Y. Perfluoroalkyl End-Capped Poly(Ethylene Oxide). Synthesis, Characterization, and Rheological Behavior in Aqueous Solution. *Macromolecules* **2003**, *36* (2), 449–457. https://doi.org/10.1021/ma011729a.
(57) Shibayama, M. Spatial Inhomogeneity and Dynamic Fluctuations of Polymer Gels. *Macromolecular Chemistry and Physics* **1998**, *199* (1), 1–30. https://doi.org/10.1002/(SICI)1521-3935(19980101)199:1%3C1::AID-MACP1%3E3.0.CO;2-M.
(58) Charbonneau, C.; Chassenieux, C.; Colombani, O.; Nicolai, T. Slow Dynamics in Transient Polyelectrolyte Hydrogels Formed by Self-Assembly of Block Copolymers. *Physical review. E, Statistical, nonlinear, and soft matter physics* **2013**, *87*, 062302. https://doi.org/10.1103/PhysRevE.87.062302.
(59) Pape, A. C. H.; Bastings, M. M. C.; Kieltyka, R. E.; Wyss, H. M.; Voets, I. K.; Meijer, E. W.; Dankers, P. Y. W. Mesoscale Characterization of Supramolecular Transient Networks Using SAXS and Rheology. *International Journal of Molecular Sciences* **2014**, *15* (1), 1096–1111. https://doi.org/10.3390/ijms15011096.
(60) Lemmers, M.; Voets, I. K.; Cohen Stuart, M. A.; der Gucht, J. van. Transient Network Topology of Interconnected Polyelectrolyte Complex Micelles. *Soft Matter* **2011**, *7* (4), 1378–1389. https://doi.org/10.1039/C0SM00767F.
(61) Baranau, V.; Tallarek, U. On the Jamming Phase Diagram for Frictionless Hard-Sphere Packings. *Soft Matter* **2014**, *10* (39), 7838–7848. https://doi.org/10.1039/C4SM01439A.
(62) Kohn, W.; Becke, A. D.; Parr, R. G. Density Functional Theory of Electronic Structure. *J. Phys. Chem.* **1996**, *100* (31), 12974–12980. https://doi.org/10.1021/jp960669l.
(63) Haworth, N. L.; Gready, J. E.; George, R. A.; Wouters, M. A. Evaluating the Stability of Disulfide Bridges in Proteins: A Torsional Potential Energy Surface for Diethyl Disulfide. *Molecular Simulation* **2007**, *33* (6), 475–485. https://doi.org/10.1080/08927020701361876.
(64) Nussinov, R.; Liu, Y.; Zhang, W.; Jang, H. Protein Conformational Ensembles in Function: Roles and Mechanisms. *RSC Chem. Biol.* **2023**, *4* (11), 850–864. https://doi.org/10.1039/D3CB00114H.
(65) Richard, J. P. Enabling Role of Ligand-Driven Conformational Changes in Enzyme Evolution. *Biochemistry* **2022**, *61* (15), 1533–1542. https://doi.org/10.1021/acs.biochem.2c00178.

## Abstract Graphic

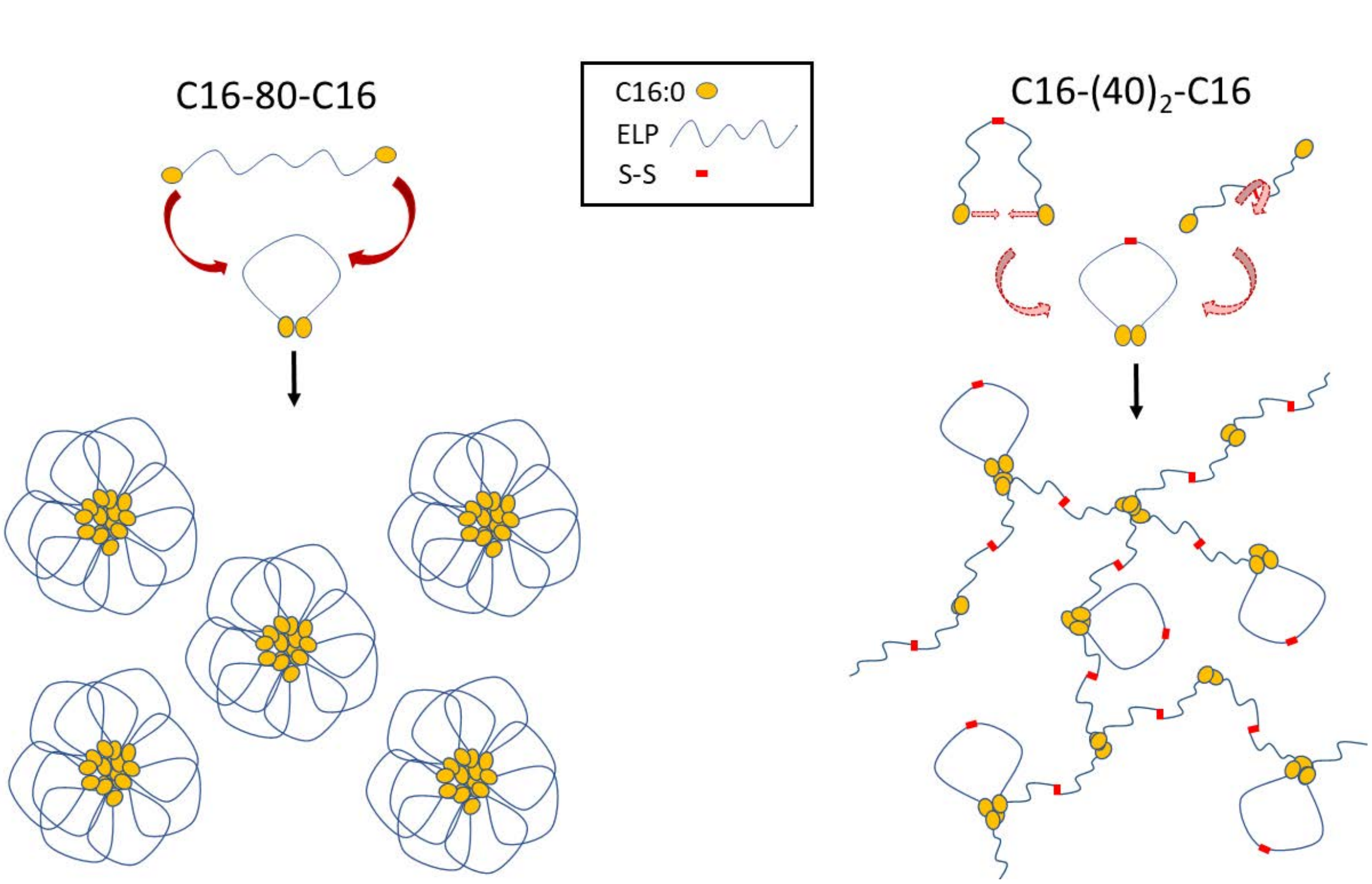

A disulfide-based biomimetic [illegible] enhances self-assembly dynamics and enables the formation of a transient polypeptide network gel at only 2 % by weight.

# Supporting Information for "A Central Disulfide Junction Drives Transient Network Formation in Elastin-Like Polypeptides, Enabling Low-Concentration Hydrogels"

Tingting Zhang,[a] Jean-François Le Meins,[a] Jean-Paul Chapel,[b] Saron Catak,[c] Guillaume Goudounet,[a] Olivier Sandre,[a] Nadia Mahmoudi,[a] Christophe Schatz*[a] and Bertrand Garbay*[a]

[a] Dr. T. Zhang, Dr. Le Meins, G. Goudounet, Dr. O. Sandre, Dr. N. Mahmoudi, Dr. C. Schatz, Prof. B. Garbay
Laboratoire de Chimie des Polymères Organiques (LCPO) UMR 5629
Bordeaux University, CNRS, Bordeaux INP
F-33600, Pessac, France

[b] Dr. J.-P. Chapel
Centre de Recherche Paul Pascal (CRPP), UMR 5031
Bordeaux University, CNRS
F-33600, Pessac, France

[c] Prof. S. Catak
Department of chemistry
Bogazici University
Bebek 34342, Istanbul, Turkey

E-mail: christophe.schatz@bordeaux-inp.fr, bertrand.garbay@bordeaux-inp.fr

## Table of contents

# S1. Materials and methods

## S1.1 Materials

Palmitic acid (Riedel-de-Haën[TM], 98%), O-(1H-6-Chlorobenzotriazole-1-yl)-1,1,3,3-tetramethyluronium hexafluorophosphate (HCTU, Novabiochem), N,N-Diisopropylethylamine (DIPEA, Sigma-Aldrich, 99.5%), N,N-Dimethylformamide (DMF, VWR chemicals, 99.9%), Diethyl ether (VWR international, 100%), Acetone (VWR international, 99%), Hydrogen peroxide (Acros organics, 35 wt.% solution in water), Phosphate Buffer Saline 10X, (PBS 10X, Euromedex) were used as received, without further purification. Ultrapure water (18 MΩ cm) was obtained by passing in-house deionized water through a Millipore Milli-Q Biocel A10 purification unit. Bacto Tryptone (Sigma), Yeast extract (Sigma), Ampicillin (Sigma) and isopropyl β-D-thiogalactopyranoside (IPTG, VWR chemicals) were used for cell culture.

## S1.2 Methods

**Reversed-phase high-performance liquid chromatography (RP-HPLC).** RP-HPLC was performed using an Ultimate 3000 system (Thermo Scientific) instrument using a hydrophobic C18 stationary phase (ZORBAX Eclipse Plus C18, 4.6 x 250 mm, 5 µm, Agilent Technologies) and a gradient of solvent A ($H_2O$ + 0.1% TFA)/solvent B (methanol + 0.1% TFA). The flow rate was 1 mL/min. 35 µL of samples (~70 µM) were injected, and the absorbance at 230 nm was recorded with a diode array detector (Thermo Scientific). Data obtained by RP-HPLC were used to calculate the percentage of purity of the ELP-C16 conjugates. This was achieved by integrating the area below the peak of the palmitoylated ELP, and by subtracting the area under the peak, if any, of the unmodified ELP.

**Cloud point temperatures ($T_{cp}$).** The $T_{cp}$ values were determined by measuring the turbidity at 350 nm between 10 and 50°C with a scan rate of 1°C·min$^{-1}$. Data were collected using a Cary 100 UV–vis spectrophotometer equipped with a multicell thermoelectric temperature controller (Agilent Technologies). The $T_{cp}$ was defined as the maximum of the first derivative of absorbance with respect to temperature.

**Dynamic light scattering (DLS).** DLS measurements were carried out using a Nano ZS instrument (Malvern, U.K.) equipped with a He-Ne laser ($\lambda_0$ = 632.8 nm) and operating at a fixed detection angle of 90°. Prior to measurement, all samples were filtered through 0.45 µm hydrophilic PVDF syringe filters (Millex, Merck Millipore) to remove dust and aggregates. Particle size distributions were obtained from the intensity autocorrelation functions using the non-negative least squares (NNLS) algorithm implemented in the instrument software.

**Static light scattering (SLS).** SLS measurements were performed using the static mode of the Nano ZS instrument, that is, by measuring the scattering intensity expressed on a absolute scale known as the derived count rate (DCR), expressed in kilo count per second (kcps). Preferentially, all analysis were performed using the same attenuator (position 11) to avoid using attenuation factor which introduce errors in the DCR value. The scattered intensities at various concentrations were analyzed according to the Rayleigh equation valid for small scatterers (d < $\lambda/20$) by plotting the reciprocal of the scattering intensity against the concentration:

$$\frac{Kc}{\Delta R} = \frac{1}{M_w} + 2A_{2,z}c$$

with $\Delta R$ the excess Rayleigh ratio, $c$ the mass concentration, $M_w$ the weight-average molar weight, $A_2$ the z-average second virial coefficient and $K$ the optical constant (contrast factor) where $K = 4\pi^2 n_0^2 (dn/dc)^2/\lambda_0^4 N_A$ with $n_0$ the refractive index of the solvent, $(dn/dc)$ the refractive index increment, $N_A$ the Avogadro number and $\lambda_0$ the vacuum wavelength of the primary beam. The excess Rayleigh ratio was obtained at various concentrations according to:

$$\Delta R = \frac{I - I_0}{I_T}\left(\frac{n_0}{n_T}\right)^2 R_T$$

with $I, I_0, I_T$ the scattered intensities of the particle dispersion, solvent and toluene measured at varying scattering angle $\theta$ and $R_T$ the Rayleigh ratio of toluene (1.355 x $10^{-5}$ $cm^{-1}$). The typical value of refractive index increment of protein ($dn/dc$ = 0.185 mL/g) was used. Due to the equilibrium between individual copolymer micelles and micelle clusters, the contribution of micelles to the total scattering intensity was evaluated from DLS analysis by considering the area of the micelle population in the intensity-weighted particle size distribution (Figure S6).

**Hydrogel formation and rheological analyses.** The triblock polymer was dissolved in milliQ water at a concentration of 5% (w/v) with stirring taking place at a temperature of 10 °C (below $T_{cp}$). Gel formation occurred after 12-24 h of continuous stirring. Rheological measurements were performed using an Anton Paar rheometer equipped with a Peltier temperature controller. Depending on the sample structure and available volume, the gel was loaded at 10 °C between either parallel steel plates or conical plates (25 mm diameter, 2° cone angle). The linear viscoelastic region was determined at an angular frequency of 1 rad/s over a strain range of 1–100%. Frequency sweep experiments were conducted from 100 rad/s down to either 0.1 or 0.01 rad/s depending on the sample. Temperature ramp experiments were performed between 4 °C and 40 °C at a heating or cooling rate of 5 °C/min. Data acquisition and analysis were carried out using Anton Paar RheoCompass™ software.

**Small-angle X-ray scattering (SAXS).** SAXS profiles were acquired using a XEUSS setup (Xenocs, Grenoble, France) comprising a microfocus copper anode source, a scatterless collimation system and a PILATUS 2D detector. This setup provided access to scattering wave vector q values ranging from 0.009 to 0.5 Å$^{-1}$. The hydrogel samples were placed in glass capillaries at room temperature, where they remained in a fluid or slightly viscous state. Measurements were performed under controlled temperature conditions using a Peltier-based temperature control system. The resulting 2D images were found to be isotropic and the data were azimuthally averaged to give the intensity scattering curve, I(q), which was corrected for the experimental background (solvent and capillary scattering). This was then converted to absolute intensity as a function of $q = (4\pi/\lambda)\sin\theta$, where $\lambda = 0.154$ nm is the wavelength of the CuKα radiation and θ is half the scattering angle. The data were fitted using SasView 6.0 software with a combination of a polydisperse spheres (Gaussian distribution) form factor yielding the micelle radius and a hard-sphere structure factor providing an effective volume fraction ϕ and a hard-sphere radius (RHS). A more sophisticated core-shell form factor model did not improve the results.[1]

**Cryo-Transmission Electron Microscopy (Cryo-TEM).** Cryo-TEM imaging was performed using a JEOL 2100 transmission electron microscope (JEOL, Japan) equipped with a LaB6 filament and operated at an accelerating voltage of 200 kV. For sample preparation, a drop of the nanoparticle suspension was applied to a holey carbon grid (Quantifoil Micro Tools GmbH, Germany), and excess liquid was removed with filter paper. The grid was then rapidly vitrified by plunging into liquid ethane, mounted on a Gatan 626 cryo-holder, and transferred to the microscope. Imaging was carried out at −180 °C using the JEOL low dose system (Minimum Dose System, MDS) to protect the thin ice film from any irradiation before imaging and reduce the irradiation during the image capture. Micrographs were acquired with an Ultrascan 2k pixel CCD camera (Gatan, USA) and analyzed using ImageJ software. Particle size distributions, mean diameters, and standard deviations were determined by applying Gaussian statistical analysis to the measured particle population.

**Computational Methodology.** A quantum-mechanical study was performed using DFT calculations[2] utilizing the Gaussian G09 software package.[3] The effect of the solvent environment on peptide conformational flexibility was taken into account by means of the self-consistent reaction field (SCRF) theory, using the polarizable continuum method (IEF-PCM)[4] in water. Geometries were optimized using (B3LYP[5,6]/6-31G(d,p)) level of theory.

## S1.3 Construction of the ELP clones

The production of bacterial clones that code for the ELP 40 and ELP 80 sequences has already been published.[7,8] To introduce a cysteine or a lysine residue at the C-terminal end of ELP40 and ELP80, respectively, we used site-directed mutagenesis. In brief, *E. coli* clones containing the pUC19 plasmids encoding ELP40 or ELP80 sequences were grown overnight at 37°C in Lysogeny Broth (LB) medium (1% Bacto Tryptone, 1% yeast extract, 0.5% NaCl) supplemented with 100 µg/L ampicillin. The plasmids were then purified using a NucleoSpin® Plasmid kit (Macherey-Nagel) according to the manufacturer's instructions. The concentration of the purified plasmids was measured by spectrophotometry at 260 nm using a Nanodrop ND 1000, and their identity was confirmed by digestion with the *Bam*HI restriction enzyme (New England Biolabs) followed by agarose gel electrophoresis. The insertion of one cysteine codon (TGT) or one lysine codon (AAA) at the C-terminal end of the ELP sequence was performed by site directed mutagenesis using the InFusion kit (Takara). Briefly, a PCR was performed using the ELP-pUC19 plasmid as the matrix and the following primers (the inserted codons are in bold):

For the insertion of the cysteine codon, forward primer ELP40-Cys: $^{5'}$GGCGTAGGT***TGT***TAAACCTCATCGCTGGATCCAAAGC$^{3'}$, and reverse primer ELP40-Cys**:** $^{5'}$TTTA***ACA***ACCTACGCCAGGTACGCC$^{3'}$.

For the insertion of the lysine codon, forward primer ELP80K: $^{5'}$CGTAGGT***AAA***TAAACCTCATCGCTGGATCCTC$^{3'}$, and Reverse primer ELP80K: $^{5'}$GTTTA***TTT***ACCTACGCCAGGTACGCC$^{3'}$.

The PCR reaction was as follows: 3 cycles of denaturation at 98°C for 10 s, hybridization at 56°C for 15 s and elongation at 72°C for 20 s, and then 30 cycles of 98°C for 10 s, hybridization at 58°C for 15 s and elongation at 72°C for 20 s. To obtain the linear PCR product from the plasmid matrix, the product of the reaction was separated in 0.8% agarose gel and the band corresponding to the PCR product was excised from the gel and purified using Monarch® DNA gel extraction kit (Biolabs). Then the linear PCR products were re-circularized using Fusion HD enzyme (Takara) and transformed into *E. coli* competent Stellar Cells from Takara [F–, endA1, supE44, thi-1, recA1, relA1, gyrA96, phoA, Φ80d lacZΔ M15, Δ (lacZYA - argF) U169, Δ (mrr - hsdRMS - mcrBC), ΔmcrA, λ–]. Clones were selected on LB agar plates containing 100 µg/mL ampicillin, and analyses by PCR using One Taq Hot Start Quick Load 2X master mix (New England Biolabs) and the primers M13 forward ($^{5'}$TGTAAAACGACGGCCAGT$^{3'}$) and M13 reverse ($^{5'}$CAGGAAACAGCTATGACC$^{3'}$). PCR was as follows: cell lysis for 3 min at 94°C, and then 35 cycles 94°C for 15 s, 50°C for 30 s and 68°C for 1 min. The PCR products were analyzed by 1.5% agarose gel electrophoresis. Plasmids from three PCR positive clones for each construction were purified with the NucleoSpin® Plasmid kit, verified by double digestion with *Bam*HI and *Nde*I, and sent for sequence determination (Eurofins Genomics)

For subcloning in the pET44a expression vector**,** the purified plasmids were digested by *Bam*HI and *Nde*I. The restriction products were separated in 1.2% agarose gel, and the bands corresponding to the inserts were purified with the Monarch® DNA gel extraction kit. In parallel, the pET44a plasmid was double digested with *Bam*HI and *Nde*I, and the linearized plasmid was purified by 0.8% agarose gel electrophoresis. Inserts and linearized pET44 plasmid were ligated in 20 µL using a 3:1 insert/vector molar ratio with the Quick ligase Kit (New England Biolabs). Then, 2.5 µL of the reaction was used to transform BLR(DE3) *E. coli* cells [F-ompT hsdSB(rB- mB-) gal lac ile dcm Δ(srl-recA)i306:Tn10 (tetR)(DE3)]. Cells were spread on LB agar plates containing 100 µg/mL ampicillin, and clones were screened by PCR using One Taq Hot Start DNA polymerase (New England Biolabs) with the primers pET-f 5'GTGAGCGGATAACAATTCCCC3' and pET-r 5'GTTATGCTAGTTATTGCTCAGCGG3'. The PCR procedure was as follows: 94°C for 3 min for cell lysis, followed by 35 cycles of 94°C for 15 s, 55°C for 30 s and 68°C for 1 min. The PCR products were analyzed by 1.5% agarose gel electrophoresis in TAE buffer. The plasmids from positive clones were purified with NucleoSpin® Plasmid kit, and their sequences verified (Eurofins genomics).

## S1.4 Production of the ELP 40 and 80

A single bacterial colony was selected and cultured overnight at 37°C in a rotary shaker at 200 rpm in 15 mL of Lysogeny Broth (LB) medium (0.4% LB Lennox, 0.1% yeast extract) containing 100 µg/mL ampicillin. This seed culture was then inoculated into 400 mL of LB medium supplemented with 1 g/L glycerol and 100 µg/mL ampicillin to obtain an initial $OD_{600nm}$ of 0.15. The culture was then grown at 37°C and 200 rpm until the $OD_{600nm}$ reached a value close to 0.6. IPTG was then added to give a final concentration of 0.5 mM, and the temperature was lowered to 25°C. After 12 h of IPTG induction, the culture was harvested by centrifugation at 7,500 g and 4 °C for 15 min, after which the cell pellet was suspended with 10 mL/g wet weight in deionized water. The samples were then incubated overnight at −80°C and slowly defrosted by incubation at 30°C. Total cell lysis was performed by sonication at 15°C with several 4-second pulses at 125 W, 9 s apart, for a total duration of 45 min. The insoluble debris was then removed by centrifugation at 11,000 g for 15 min at 4°C. The cleared lysate (soluble fraction) was then subjected to three successive inverse transition cycles (ITC).[9] In brief, the ELPs were precipitated at 30°C in the presence of salts, then pelleted by centrifugation for 15 min at 7,000 g and 30°C. After the supernatant containing soluble proteins was eliminated and the pellet was re-suspended in cold water, the ELPs were re-dissolved, and insoluble proteins from E. coli were eliminated by centrifugation for 15 min at 11,000 g and 4 °C. The supernatant containing the ELPs was then processed through the second and third ITC cycles. Before each 'warm spin', NaCl was added to reach final concentrations of 1.5 M, 1 M and 0.5 M for the first, second and third cycles, respectively. Finally, the purified ELPs were dialyzed against ultrapure water and lyophilized.

## S1.5 Dimerization of ELP 40

To trigger the formation of intermolecular disulfide bridges, we used mildly oxidative conditions. We dissolved 80 mg (4.6 µM) of ELP 40 in 4 mL of Tris buffer (Tris 10 mM, pH 8.8) at 4 °C. Then, 1.5 ml of 0.05% $H_2O_2$ (5 equivalents) was added and the mixture was incubated for 30 min at 25°C. The oxidized ELP was dialyzed with a 1 kDa cut-off membrane at 10 °C for two days against water, and then lyophilized. SDS-PAGE was performed to determine the quantities of ELP 40 monomers and ELP $(40)_2$ dimers present.

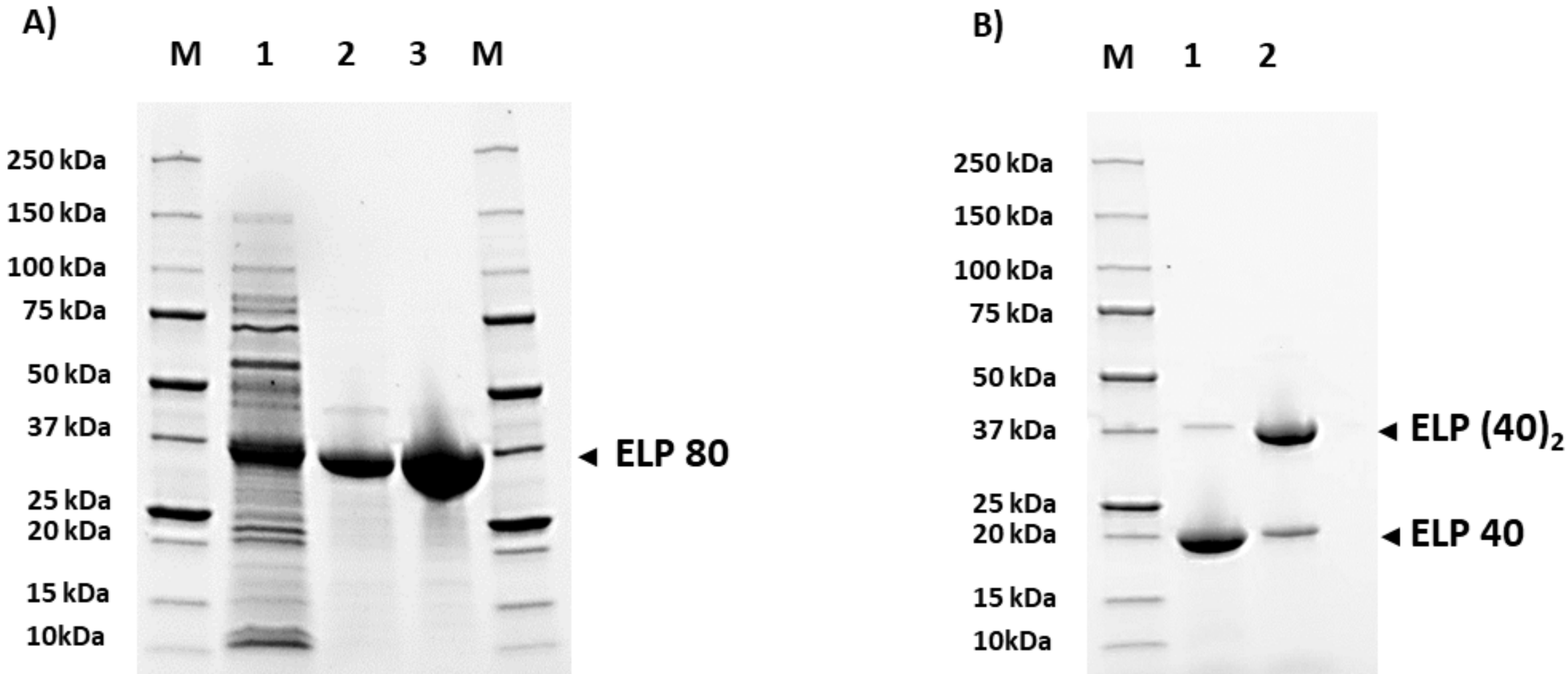

**Figure S1. SDS-PAGE analysis of the purified ELP**. **A)** Purification of the recombinant ELP 80 protein. Lane M: protein marker (Precision Plus Polypeptide Standard). Lane 1: *E. coli* soluble fraction. Lane 2: Purified ELP 80 after the first ITC cycle. Lane 3: Purified ELP 80 after the third ITC cycle. Some traces of low-molecular-weight contaminants were detected in the purified ELP 80 after the third ITC cycle (lane 3), but these represent less than 2 % of the total protein content as estimated by densitometry analysis of the gel. **B)** Lane M: protein marker (Precision Plus Polypeptide Standard). Lane 1: purified ELP 40 after three ITC cycles. Lane 2: oxidized ELP 40. We obtained highly pure ELP 40 (lane 1), which was predominantly present in monomeric form; the quantity of dimer was low, accounting for less than 3% after densitometric analysis of the gel. This was expected because the presence of reductases in the cytoplasm of *E. coli* prevents the stability of disulfide bridges.[10] Following oxidation, the proportion of dimers in the ELP 40 solution increased, accounting for 95% of the sample (lane 2).

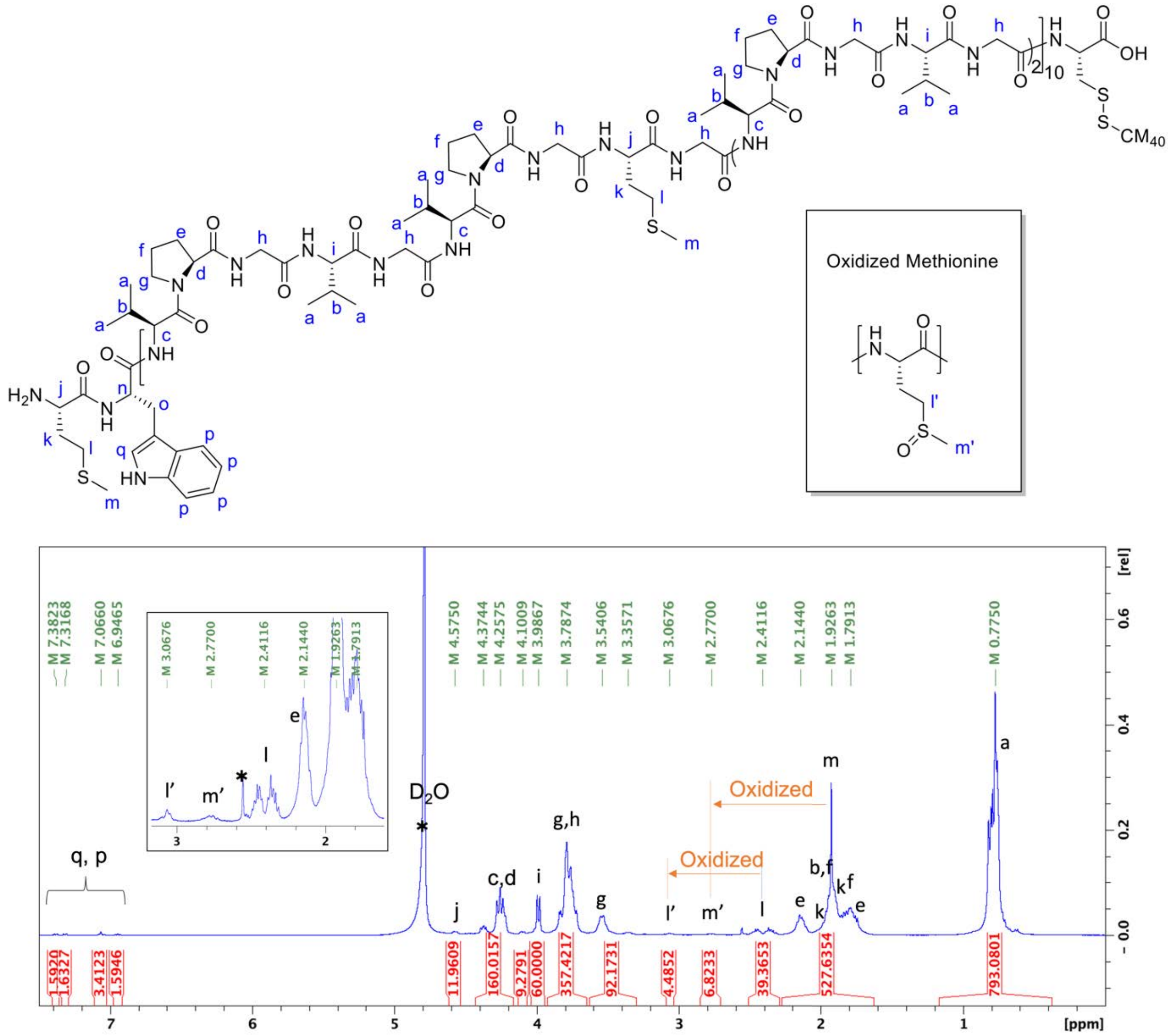

**Figure S2. $^1$H NMR analysis of ELP $(40)_2$ in $D_2O$.** We verified that oxidized ELP40 did not contain a significant amount of sulfoxide residues. Oxidation of methionine residues in ELPs can lead to sulfoxide formation, which is known to increase the cloud point temperature.[8] Upon methionine oxidation, the NMR signals of protons adjacent to the sulfur atom shift downfield: from 2.41 ppm for γCH2 Met (l) to 3.06 ppm (l'), and from 1.92 ppm for $\varepsilon CH_3$ Met (m) to 2.77 ppm (m'). Integration of these peaks allowed us to determine the sulfoxide/methionine ratio in oxidized ELP, which was found to be ~10%, corresponding to 2 sulfoxides and 20 unoxidized methionines in ELP$(40)_2$.

## S1.6 Acylation of the ELPs

The palmitoylation of ELP was performed as previously described.[11] The first step was to activate the fatty acid. A solution of palmitic acid was prepared at 3 mM in anhydrous DMF and stirred for 1 h at room temperature (RT). Carboxylic acid activation was achieved by adding 7.4 mg (18 µmol) of HCTU and 4.6 µL (26 µmol) of DIPEA to the solution containing 18 µmol of palmitic acid, and then performing the reaction in the dark for 15 min. The second step involved conjugating the activated fatty acid with the primary amine groups at the ends of ELP 80 and ELP $(40)_2$. For this, 3 µmol of the activated fatty acid was reacted with 1 µmol of ELP in the dark for 12 h. The reaction product was then precipitated using 10 volumes of acetone, collected by centrifugation for 5 min at 2,500 g and 20 °C, and re-suspended in DMF. Following a second precipitation step involving 10 volumes of diethyl ether, the resulting pellets were recovered by centrifugation and washed three times with 10 volumes of diethyl ether. After drying for 2 h at room temperature, the product was dissolved in cold water and centrifuged at 2,500 g for 10 min at 4 °C to remove insoluble impurities. The clear supernatant was dialyzed (1 kDa Spectra/Por® membrane) against ultrapure water for 2 days at 4 °C. The product was then lyophilized and stored at -20 °C.

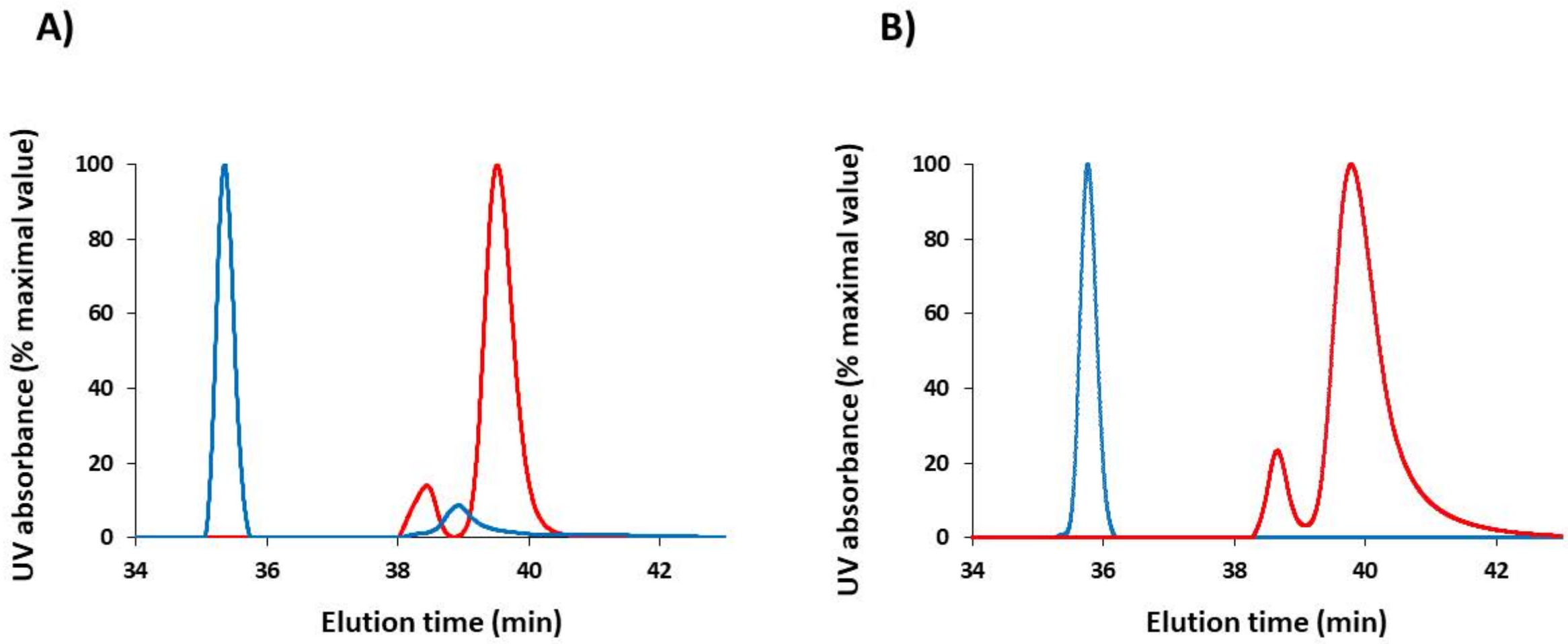

**Figure S3. RP-HPLC traces of ELPs before and after acylation. A)** Traces for ELP 80 (blue line) and palmitoylated ELP 80 (red line). Prior to the acylation of ELP 80, a major peak was observed at 35.4 min and a smaller peak at 38.9 min. This secondary peak could correspond to small amounts of ELP 80 dimers, although no dimers were detected in the SDS-PAGE analysis (lane 3, Fig. S1A), suggesting that they are not covalently linked. Following palmitoylation, the initial peaks disappeared, and two new peaks appeared at 38.4 and 39.5 min. The larger peak, which accounts for 90% of the product, corresponds to the expected C16-80-C16 triblock, while the smaller peak represents ELP 80 that has been palmitoylated at either the N- or C-terminus only. **B)** Traces for ELP $(40)_2$ (blue line) and palmitoylated $(40)_2$ (red line). After palmitoylation, 92% of the product was present as the triblock (retention time of 39.8 min), and 8% corresponded to mono-palmitoylated ELP**.**

# S2. Supplementary characterization data

## S2.1 Optical microscopy analysis of ELP solution

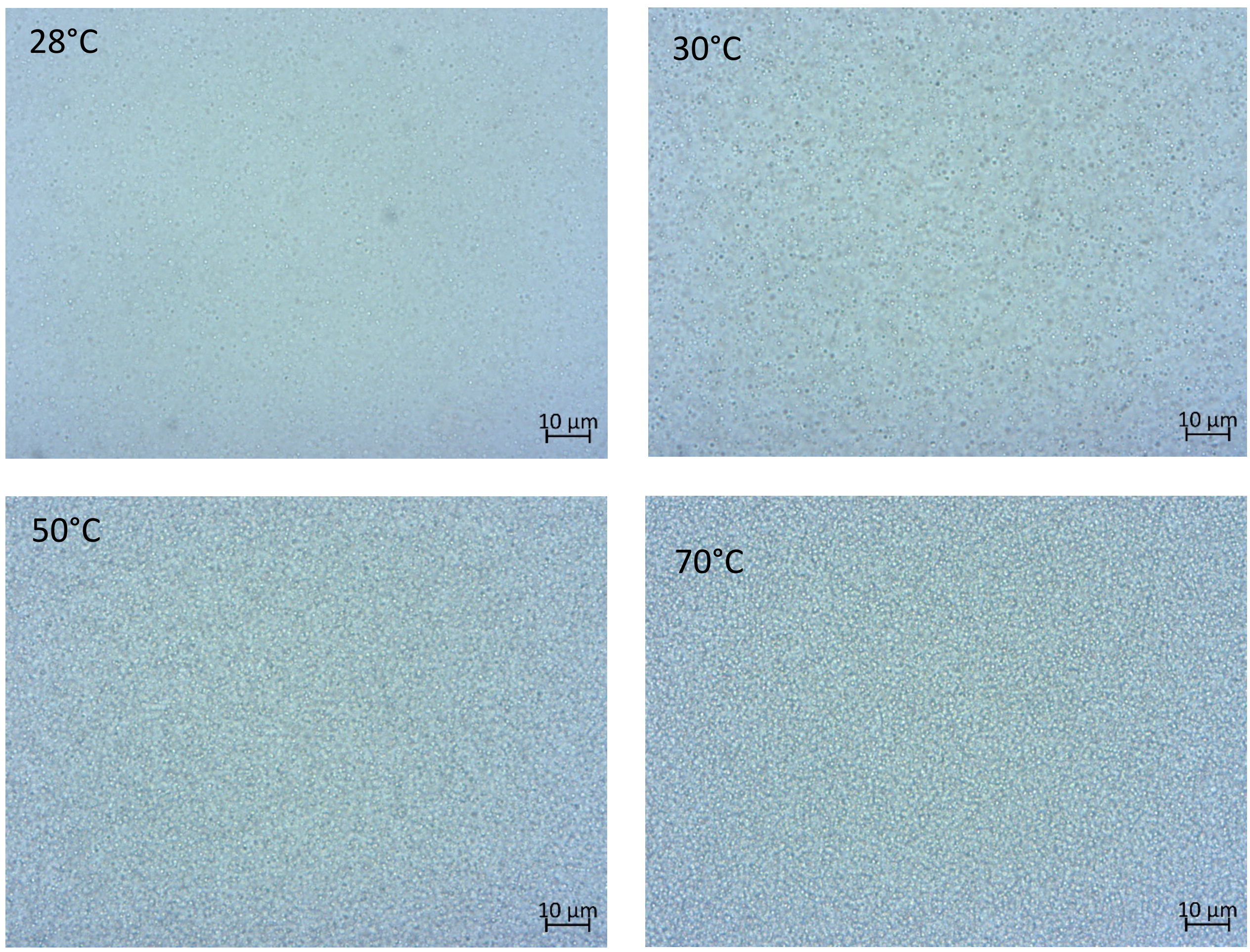

**Figure S4.** Optical microscopy of a 2.5 mg/mL ELP 80 solution at various temperatures, showing the liquid–liquid phase transition associated with thermally triggered coacervation. Upon heating, droplets become smaller, denser, and more numerous.

## S2.2 Light scattering analysis of acylated ELPs

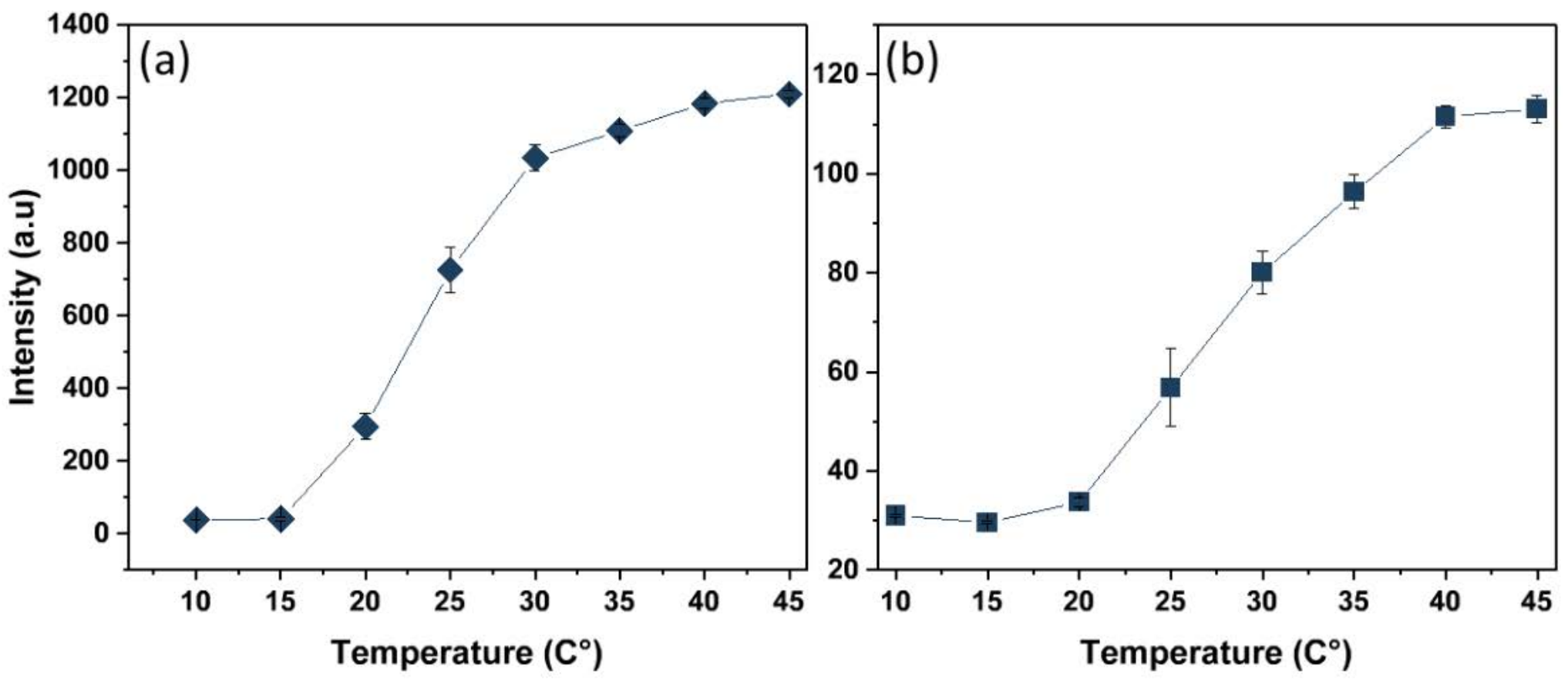

**Figure S5** Scattered intensities of C16-80-C16 (a) and C16-$(40)_2$-C16 (b) at c = 2.5 mg/mL, between 10°C and 45°C.

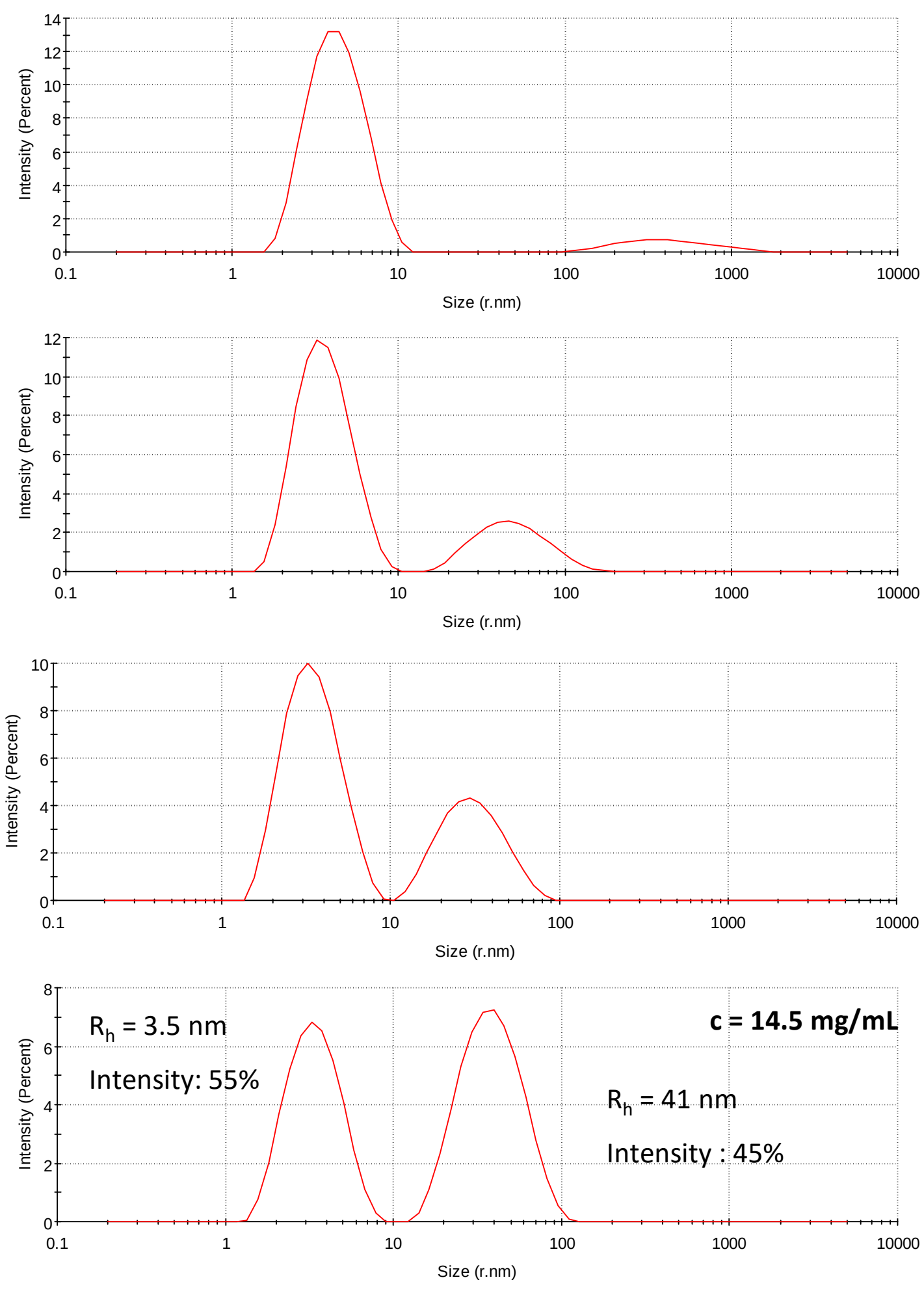

**Figure S6.** Intensity-weighted hydrodynamic radii ($R_h$) of C16-$(40)_2$-C16 at various concentrations at 18°C. Note that the micelle size (left-hand peak) shows little dependence on concentration, consistent with a closed association mechanism.

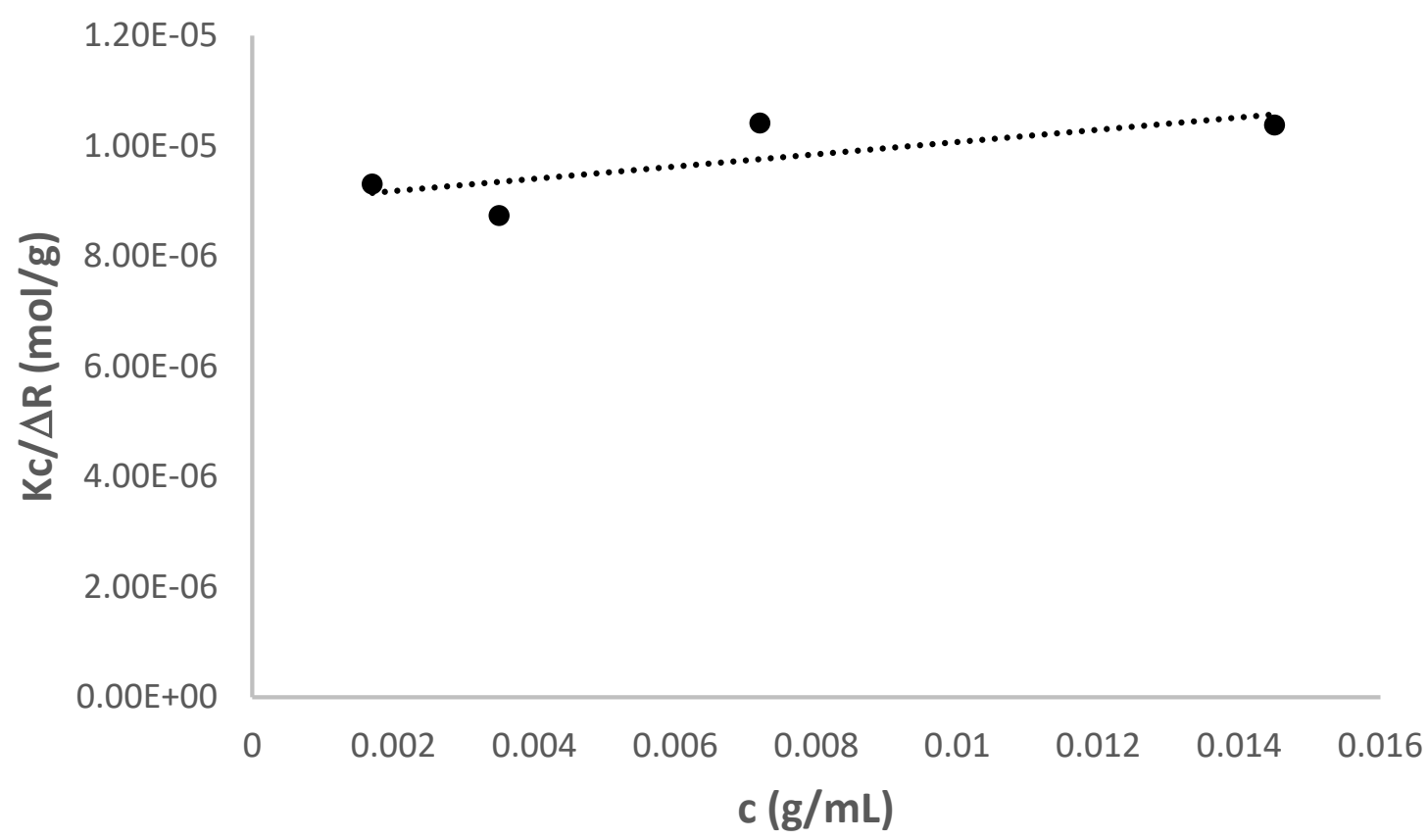

**Figure S7.** Static light scattering analysis of C16-$(40)_2$-C16 at various concentrations at 18°C. The inverse of the scattering intensity, expressed as the reciprocal of the Rayleigh ratio, is plotted as a function of concentration (Debye plot). The contribution of micelles to the total scattering intensity was evaluated from DLS analysis (Figure S6).

## S2.3 Small-angle X ray scattering analysis of ELPs and acylated ELPs

- **Scattering by the unassembled, non-acylated ELP chains**

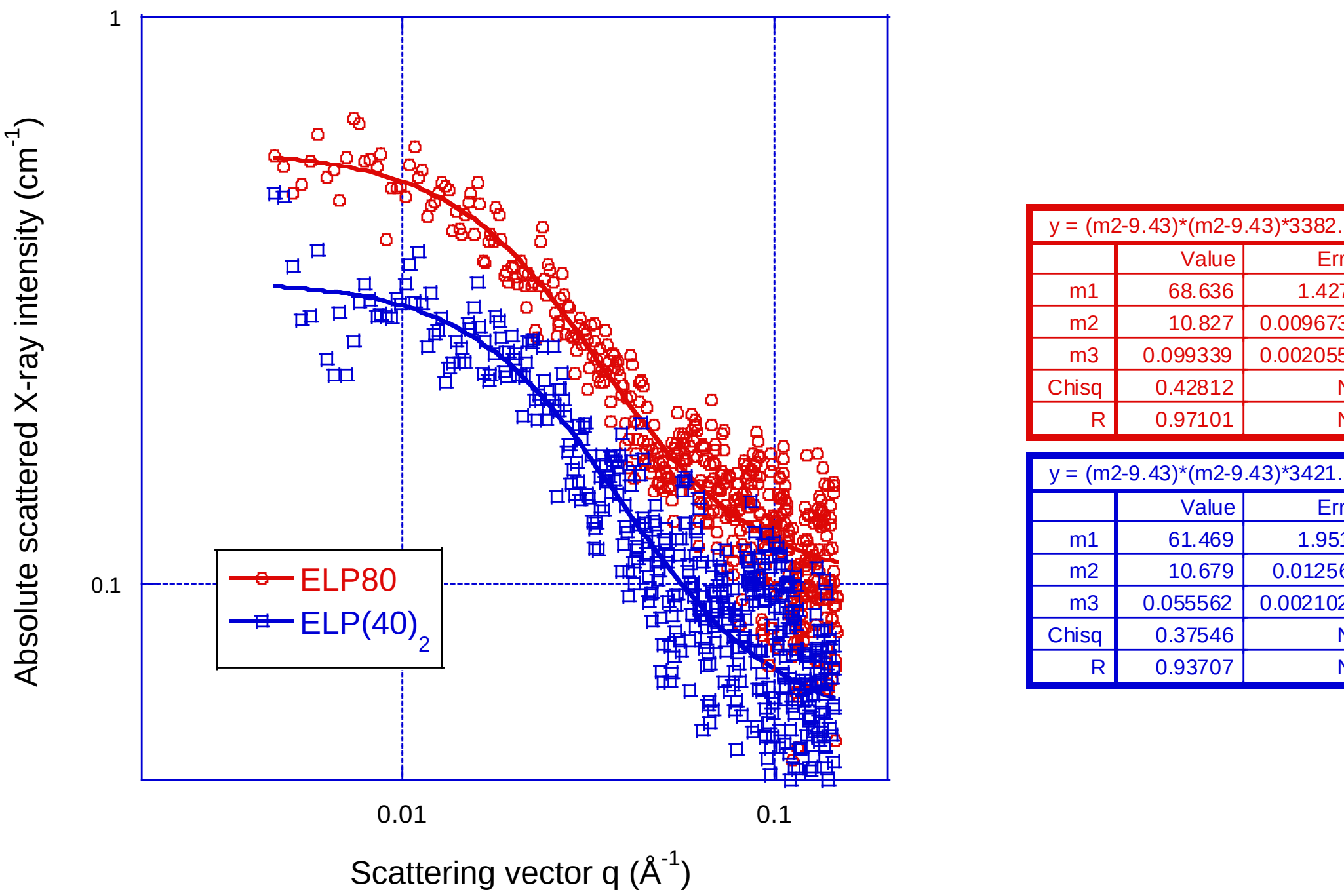

**Figure S8**. Small-angle X-ray scattering (SAXS) analysis of ELP80 and ELP(40)$_2$ solutions at 0.061 and 0.045 g·cm$^{-3}$, respectively, measured at 20 °C.

The data plotted in Figure S8 were fitted using a Debye function for the form factor P(q):

$$I(q) = \Delta\rho^2 \frac{c}{d^2\, N_{\mathrm{A}}}\, M_{\mathrm{w}}\, P(q) + I_{\mathrm{bkg}} \text{ with } P_{\mathrm{Debye}}(q, R_{\mathrm{G}}) = \frac{2}{(q^2R_{\mathrm{G}}^2)^2}\left(\exp(-q^2R_{\mathrm{G}}^2) + q^2R_{\mathrm{G}}^2 - 1\right)$$

In this expression, $\Delta\rho$ is the X-ray scattering length density (SLD) contrast between the solvent (water) and the macromolecules of molar mass $M_{\mathrm{w}}$, mass density *d* and concentration *c* (in g·cm$^{-3}$) and radius of gyration $R_{\mathrm{G}}$, $N_A$ stands for the Avogadro number, and $I_{bkg}$ for the incoherent background. The SLD can be estimated theoretically from the atomic formulas of the non-acylated ELP chains, respectively $C_{1558}H_{2530}N_{408}O_{407}S_{24}$ for ELP(40)$_2$ (dM40C, $M_w$=34218 g·mol$^{-1}$) and $C_{1542}H_{2513}N_{405}O_{404}S_{21}$ for ELP80 (M80K, $M_w$=33823 g·mol$^{-1}$). This classical calculation uses the atomic numbers of every constituting atom, each electron in their atomic shell contributing to the X-ray scattering length by $b_e^{XR}$=2.818·10$^{-13}$ cm. If we infer a mean value peptide mass density *d*=1.18 g·cm$^{-3}$, we obtain a theoretical SLD of 10.75·10$^{10}$ cm$^{-2}$ for both ELPs, as compared to 9.43·10$^{10}$ cm$^{-2}$ for water, creating the contrast $\Delta\rho$ of X-ray scattering. We used *Kaleidagraph 4.5* solver to fit the experimental SAXS curves of ELP solutions at preset experimental concentrations respectively *c*=0.045 g·cm$^{-3}$ for ELP(40)$_2$ and *c*=0.061 g·cm$^{-3}$ for ELP80, while constraining also the molar masses and mass density (values above), leaving 3 floating parameters: the $R_{\mathrm{G}}$ and SLD of the chains, and the background intensity of solvent (respectively *m*1, *m*2 and *m*3 fitting parameters). This procedure leads to SLD and gyration radius values of respectively

10.7·10[10] cm$^{-2}$ and $R_G$= 61.5 Å for the ELP (40)$_2$, 10.8·10[10] cm$^{-2}$ and $R_G$= 68.6 Å for ELP 80. The fitted contrast stays close to the theoretical value 10.75·10[10] cm$^{-2}$ calculated for both ELPs, confirming the inferred mean mass density of 1.18 g·cm$^{-3}$ (this value being higher than 1 being ascribed to internal H-bonds of peptides). Regarding the fitted gyration radii, they are both close to the power law estimate for ELP unimers from the work by Garanger *et al.*[12], $R_G = 0.36 \times M_w^{0.5}$ that gives $R_G = 66.4$ Å for mean $M_w$~34 kg·mol$^{-1}$.

- **Scattering of self-assembled acylated ELPs**

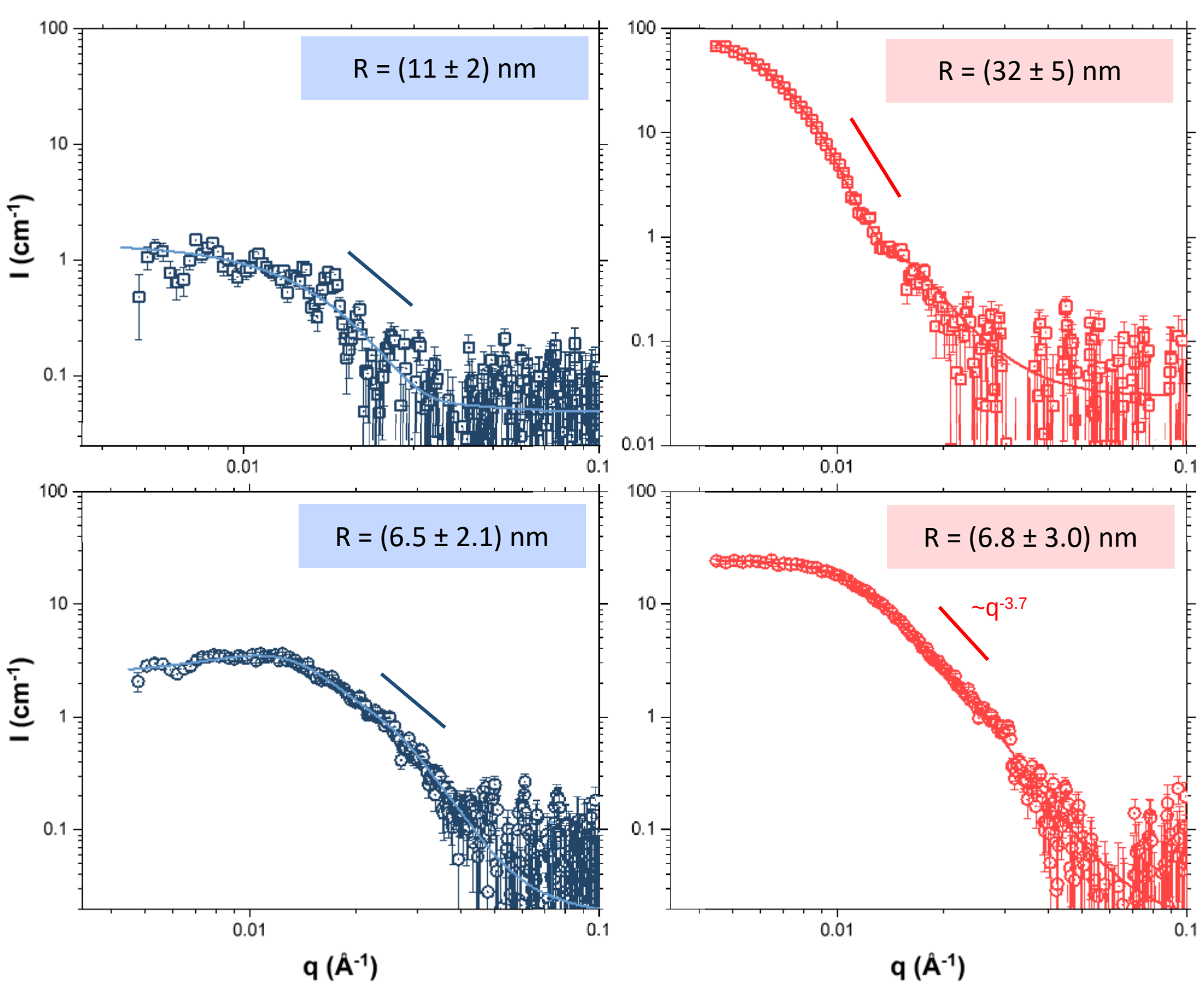

**Figure S9**. Small-angle X-ray analysis of C16-80-C16 and C16-(40)$_2$-C16 solutions (10 mg/mL) measured at 10°C and 30°C. Fits were performed using SasView.

The data in Figure S9 were first fitted using a generalized Gaussian chain model to assess the presence of unimers. As shown in the figure, these fits were unsuccessful, as the q-dependence in the mid-q range exceeded 3. A core-shell form factor was subsequently employed, based on the suspicion that micelles were present. Unfortunately, the more dilute shell was not detectable by SAXS. The C16-80-C16 data were ultimately fitted using a simple polydisperse sphere form factor P(q, R) with a Log-Normal distribution of radius and S(q)=1.

$$I(q) = \Delta\rho^2 \frac{c}{d^2\, N_A} M_w\, P(q)\, .S(q)$$

with $$P(q) = \int_0^\infty \left[\frac{3\sin(qR) - qR\cos(qR)}{(qR)^3}\right]^2 \cdot \frac{1}{R\sigma\sqrt{2\pi}} \exp\left(-\frac{(\ln R - \mu)^2}{2\sigma^2}\right) dR$$

To properly describe the C16-$(40)_2$-C16 data, a hard-sphere structure factor *S(q)* was, however, additionally included to account for the presence of clusters, which were also well-detected by DLS (Figure 2 in the main text). This is not surprising, as the measurements were performed at 1% w/w, a concentration close to the gelation threshold of 2.5% w/w. S(q) uses the Percus-Yevick closure relationship where the interparticle potential U(r) is U(r)= $\infty$ for $< 2R$ or 0 for $r \geq 2R$ (see SASview for details).

The SAXS analysis of the acylated ELPs highlights three main features: (i) the markedly smaller size of the C16-$(40)_2$–C16 assemblies compared with C16-80-C16; (ii) their weak dependence on temperature, in agreement with DLS; and (iii) characteristic dimensions that are slightly larger than the hydrodynamic radius of the small DLS population (Figure 2), likely reflecting the coexistence of small associative structures and larger clusters in solution.

## S2.4 Gel characterization

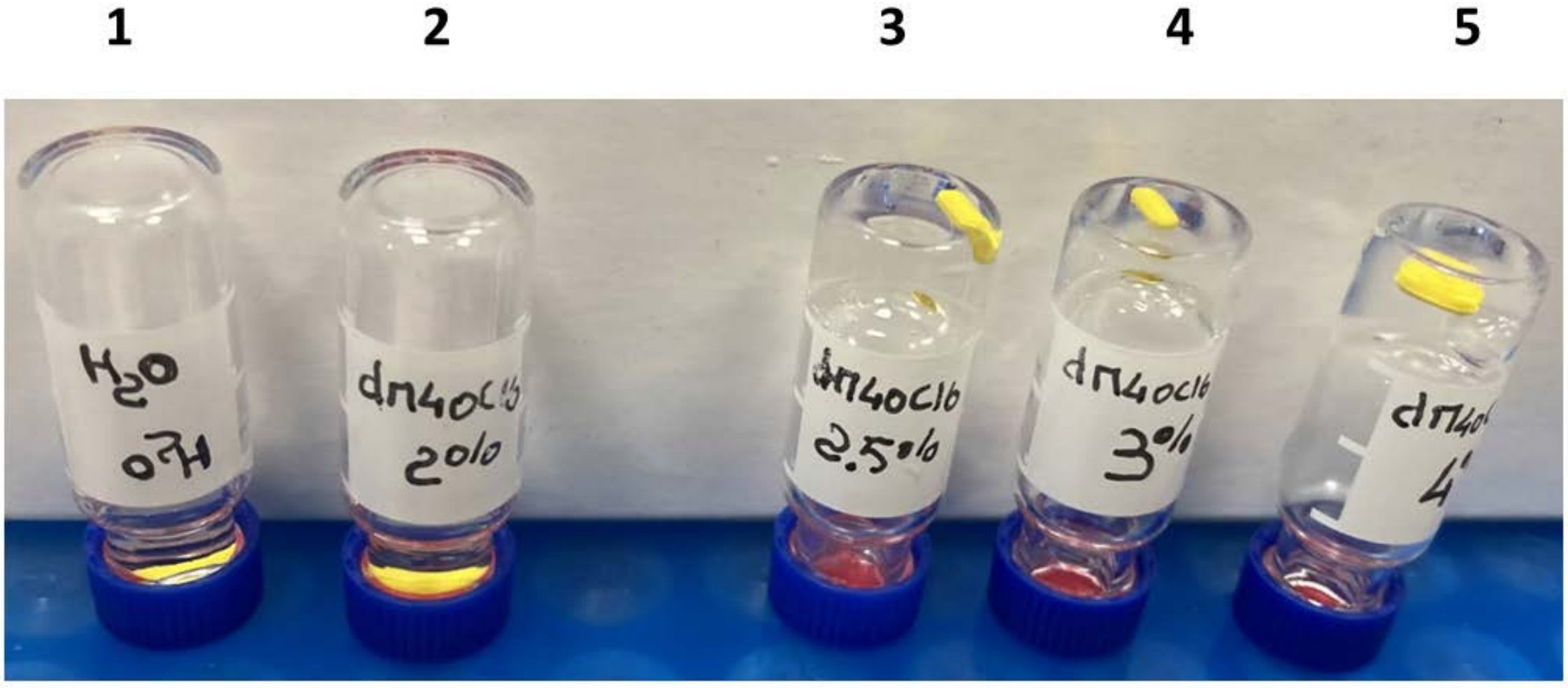

**Figure S10. Solution behavior of C16-(40)$_2$-C16 at various concentrations.** Vials containing water (1) and C16-(40)$_2$-C16 solutions at concentrations of 2 wt.% (2), 2.5 wt.% (3), 3 wt.% (4), and 4 wt.% (5) were incubated under stirring at 10 °C for 10 min using a yellow magnetic stir bar. After incubation, the vials were removed from the refrigerator, inverted, and visually inspected to assess gel formation.

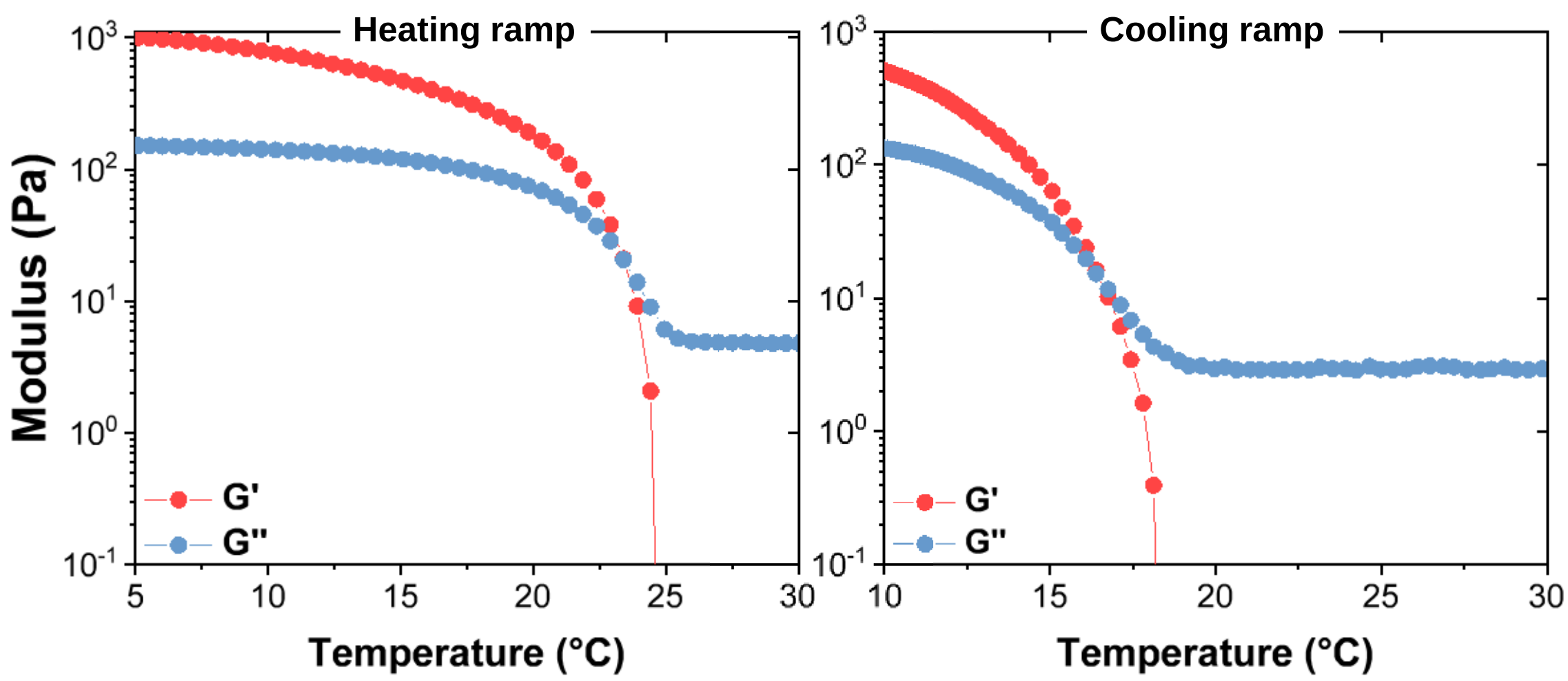

**Figure S11.** Variations of the storage (G′) and loss (G′′) moduli of C16-(40)$_2$-C16 (5 wt.% in water) as a function of temperature during heating (left) and cooling (right) ramps ($\omega$ = 40 rad s$^{-1}$, $\gamma$ = 1%, 4 °C min$^{-1}$).

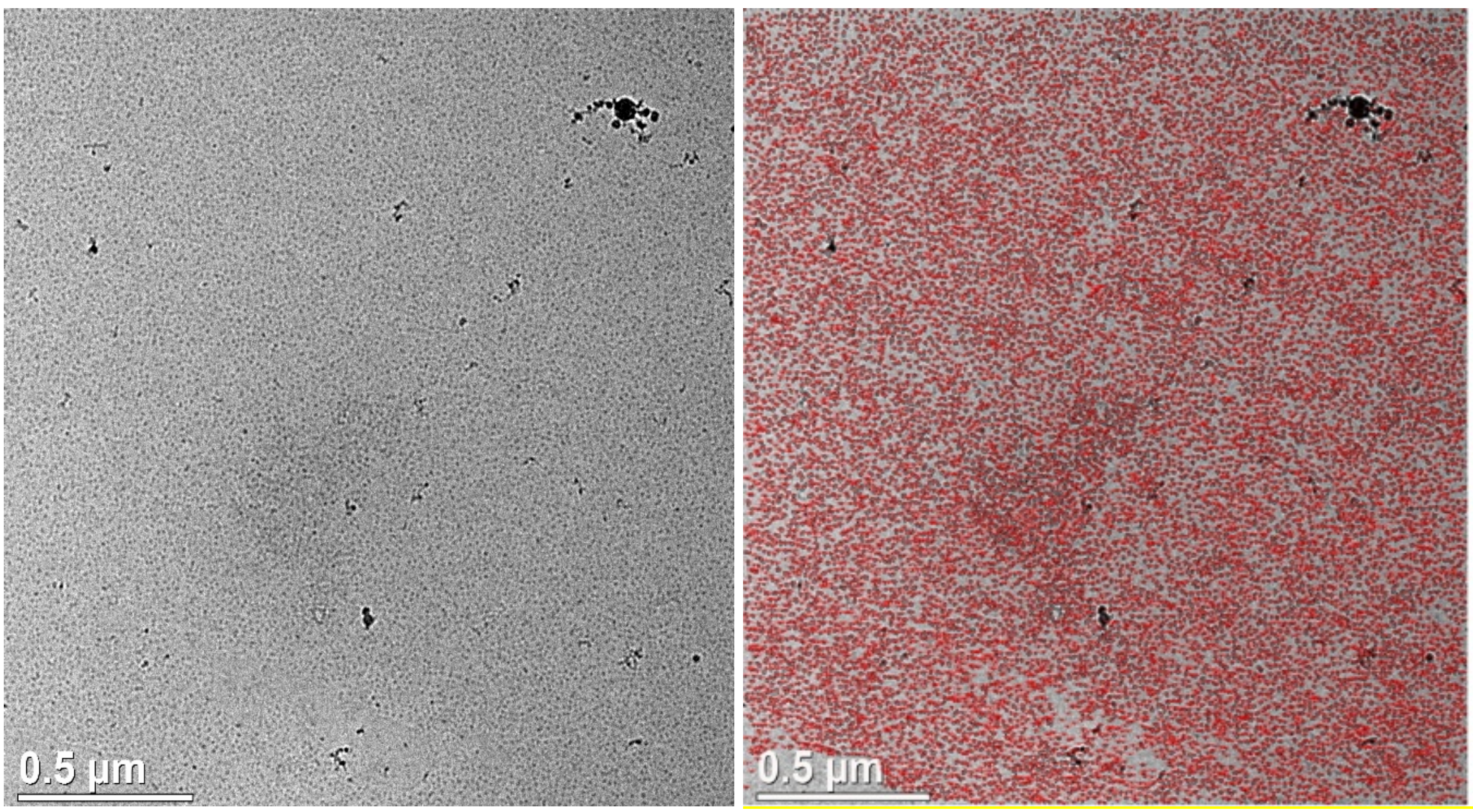

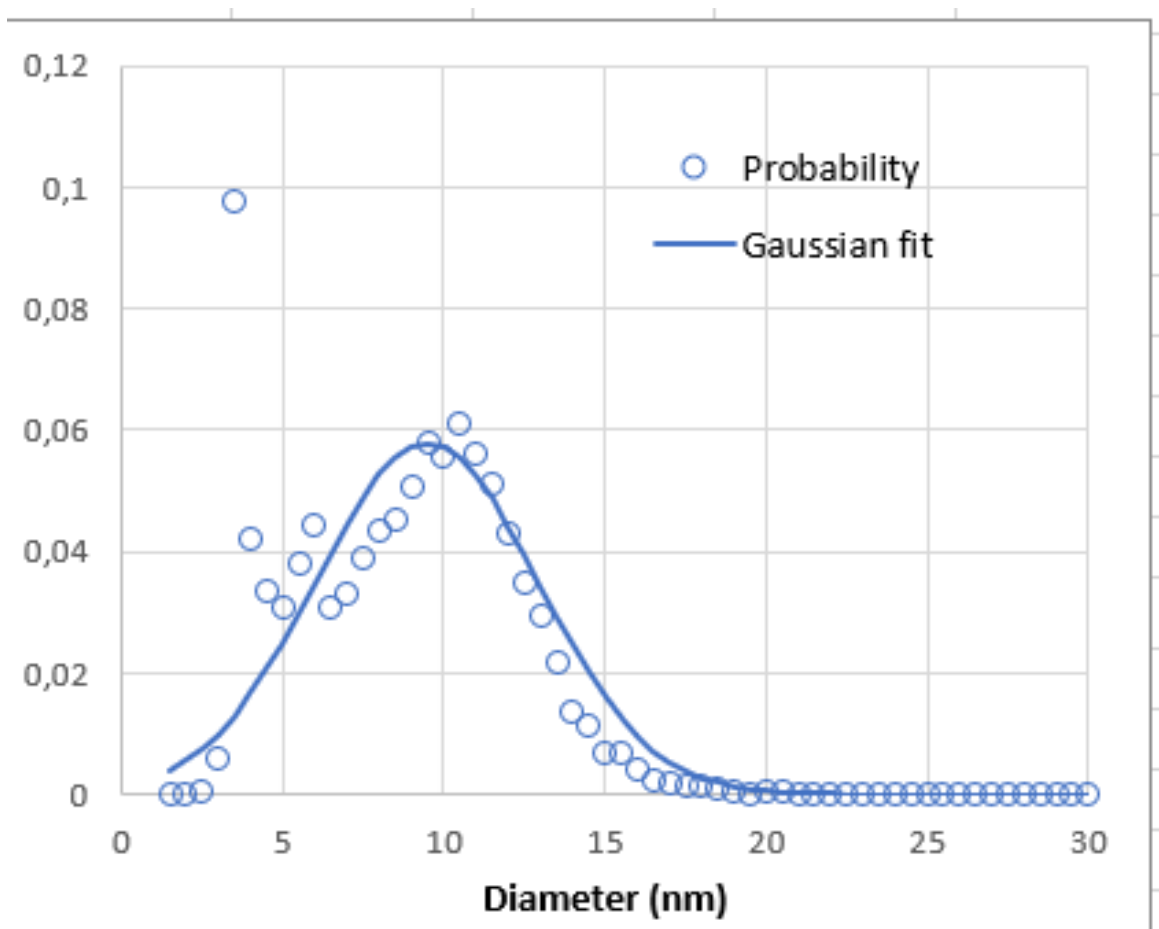

| | |
|---|---|
| Minimum | 1,99 |
| Maximum | 66,81 |
| Mean | 8,44 |
| Std dev | 3,42 |
| Nombre | 15930 |
| Intégrale | 1 |
| Total error | 0,0012033 |
| Intégale fit | 1 |
| | |
| loi de Gauss: | |
| $D_0$ | 9,48 |
| sig | 3,45 |

**Figure S12. Cryo-TEM analysis of the C16-(40)$_2$-C16 hydrogel**. Cryo-TEM image of the C16-(40)$_2$-C16 hydrogel prepared at 5 wt.% and 10 °C, with statistical analysis of size distribution from ~16,000 particles. The diameter distribution is fitted with a Gaussian function; slight deviations at small diameters are likely due to image segmentation in ImageJ. Analysis gives an average particle diameter of ~10 nm with a standard deviation of 3.4 nm.

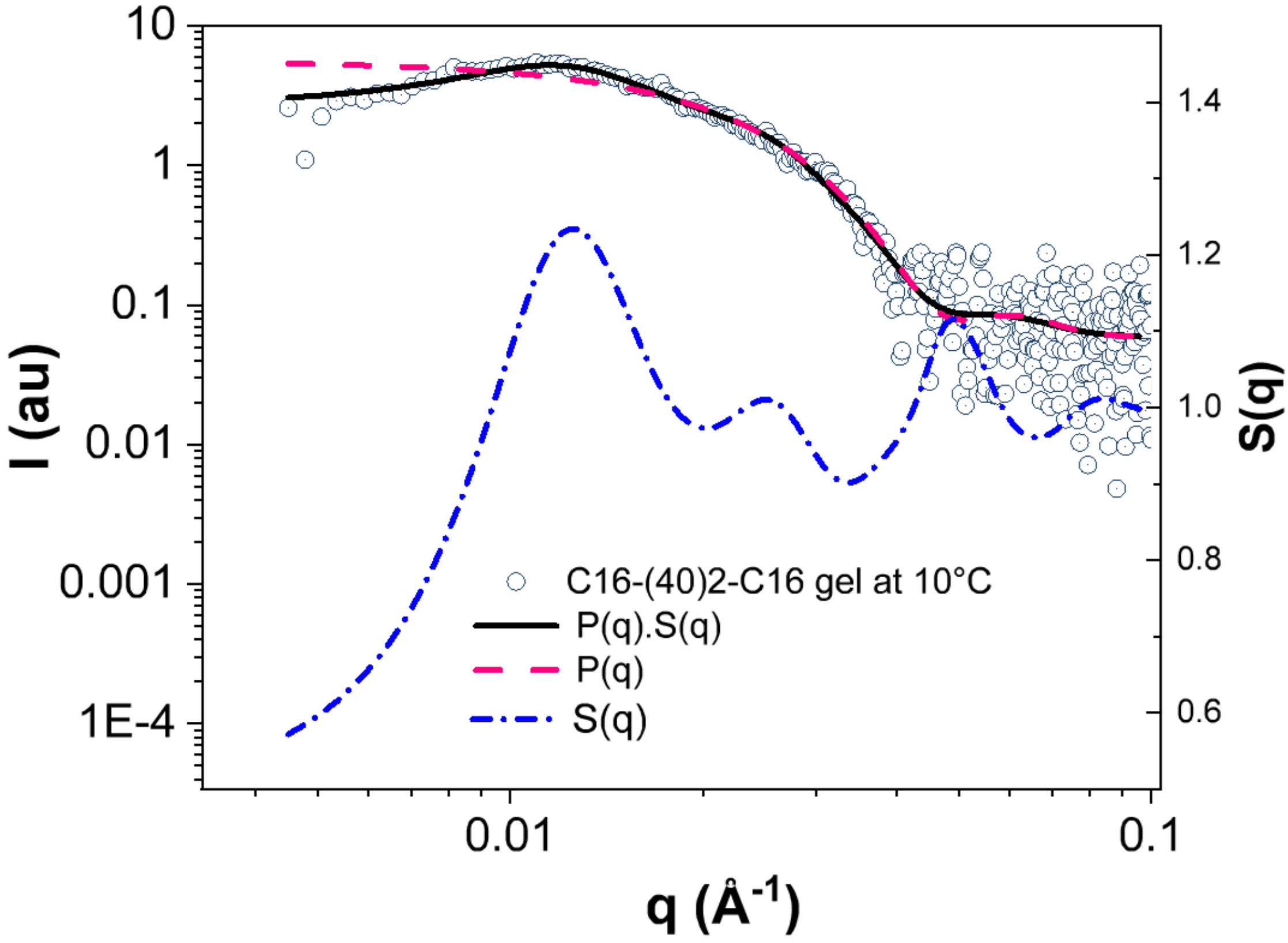

**Figure S13**. SAXS analysis of the C16-$(40)_2$–C16 gel at 10 °C at a concentration of 5 % by weight.

The contributions from the shape factor and the structure factor of C16-$(40)_2$–C16 gel at 10 °C are shown separately in Figure S13. In the low-q region, deviations between the total fit and P(q) reveal the influence of the structure factor (interactions and correlations), which exhibits a distinct peak. At high q values, where S(q) approaches 1, the total fit overlaps with P(q), confirming that particle shape dominates in this regime. The position of the correlation peak q* can be estimated from the first peak of S(q). The average center-to center distance between objects can then be estimated to $L \sim 2\pi/q^*$.

## S3. References

(1) Lemmers, M.; Voets, I. K.; Cohen Stuart, M. A.; der Gucht, J. van. Transient Network Topology of Interconnected Polyelectrolyte Complex Micelles. *Soft Matter* **2011**, *7* (4), 1378–1389. https://doi.org/10.1039/C0SM00767F.

(2) Kohn, W.; Becke, A. D.; Parr, R. G. Density Functional Theory of Electronic Structure. *J. Phys. Chem.* **1996**, *100* (31), 12974–12980. https://doi.org/10.1021/jp960669l.

(3) Frisch, M. J. Gaussian 09, Revision D.01. 2013.

(4) Tomasi, J.; Mennucci, B.; Cammi, R. Quantum Mechanical Continuum Solvation Models. *Chem. Rev.* **2005**, *105* (8), 2999–3094. https://doi.org/10.1021/cr9904009.

(5) Becke, A. D. A New Mixing of Hartree–Fock and Local Density-functional Theories. *J. Chem. Phys.* **1993**, *98* (2), 1372–1377. https://doi.org/10.1063/1.464304.

(6) Lee, C.; Yang, W.; Parr, R. G. Development of the Colle-Salvetti Correlation-Energy Formula into a Functional of the Electron Density. *Phys. Rev. B* **1988**, *37* (2), 785–789. https://doi.org/10.1103/PhysRevB.37.785.

(7) Dai, M.; Georgilis, E.; Goudounet, G.; Garbay, B.; Pille, J.; Van Hest, J. C. M.; Schultze, X.; Garanger, E.; Lecommandoux, S. Refining the Design of Diblock Elastin-Like Polypeptides for Self-Assembly into Nanoparticles. *Polymers* **2021**, *13* (9), 1470. https://doi.org/10.3390/polym13091470.

(8) Petitdemange, R.; Garanger, E.; Bataille, L.; Dieryck, W.; Bathany, K.; Garbay, B.; Deming, T. J.; Lecommandoux, S. Selective Tuning of Elastin-like Polypeptide Properties via Methionine Oxidation. *Biomacromolecules* **2017**, *18* (2), 544–550. https://doi.org/10.1021/acs.biomac.6b01696.

(9) Meyer, D. E.; Chilkoti, A. Purification of Recombinant Proteins by Fusion with Thermally-Responsive Polypeptides. *Nat. Biotechnol.* **1999**, *17* (11), 1112–1115. https://doi.org/10.1038/15100.

(10) Rietsch, A.; Beckwith, J. THE GENETICS OF DISULFIDE BOND METABOLISM. *Annu. Rev. Genet.* **1998**, *32* (1), 163–184. https://doi.org/10.1146/annurev.genet.32.1.163.

(11) Zhang, T.; Peruch, F.; Weber, A.; Bathany, K.; Fauquignon, M.; Mutschler, A.; Schatz, C.; Garbay, B. Solution Behavior and Encapsulation Properties of Fatty Acid–Elastin-like Polypeptide Conjugates. *RSC Adv.* **2023**, *13* (3), 2190–2201. https://doi.org/10.1039/D2RA06603C.

(12) Garanger, E.; MacEwan, S. R.; Sandre, O.; Brûlet, A.; Bataille, L.; Chilkoti, A.; Lecommandoux, S. Structural Evolution of a Stimulus-Responsive Diblock Polypeptide Micelle by Temperature Tunable Compaction of Its Core. *Macromolecules* **2015**, *48* (18), 6617–6627. https://doi.org/10.1021/acs.macromol.5b01371. (the authors (among which some of us) forgot to consider the S-S bridges between the terminal cysteines, therefore they had a mistake by a factor 2 on the true molar mass. Thus, the pre-factor of the law needs to be corrected to $R_G = 0.36$ Å $\times$ $M^{0.486}$)